\documentclass[11pt,a4paper]{article}
\pdfoutput=1 

\usepackage{jheppub}
\usepackage[svgnames,x11names]{xcolor} 

\usepackage[utf8]{inputenc}
\usepackage{dsfont}
\usepackage{amsfonts}
\usepackage{amsmath}
\usepackage{amsthm}
\usepackage{mathabx}
\usepackage{esint}
\usepackage{graphicx}

\newcommand{\ii}{\textrm{i}}
\newcommand{\dd}{\textrm{d}}
\newcommand{\e}{\textrm{e}}

\newcommand{\ga}{\alpha}
\newcommand{\gb}{\beta}
\renewcommand{\gg}{\gamma}
\newcommand{\gd}{\delta}
\newcommand{\gep}{\epsilon}
\newcommand{\gz}{\zeta}

\newcommand{\gth}{\theta}

\newcommand{\gl}{\lambda}

\newcommand{\gr}{\rho}

\newcommand{\Gg}{\Gamma}
\newcommand{\Gd}{\Delta}

\newcommand{\Gl}{\Lambda}

\newcommand{\Gw}{\Omega}
\newcommand{\Y}{\Upsilon}

\newcommand{\cA}{\mathcal A}
\newcommand{\cB}{\mathcal B}
\newcommand{\cC}{\mathcal C}

\newcommand{\cK}{\mathcal K}

\newcommand{\cN}{\mathcal N}
\newcommand{\cO}{\mathcal O}

\newcommand{\cR}{\mathcal R}

\renewcommand{\Re}{\mbox{Re~}}

\newcommand{\Tr}{\mbox{Tr}}

\newcommand{\be}{\begin{equation}}
\newcommand{\bea}{\begin{eqnarray}}
\newcommand{\ee}{\end{equation}}
\newcommand{\eea}{\end{eqnarray}}
\renewcommand{\d}{\partial}

\newcommand{\half}{\frac{1}{2}}

\newcommand{\ov}[1]{\frac{1}{#1}}
\newcommand{\bb}{~\scalebox{0.6}[1]{$>\hspace{-4pt}>$}~}

\title{Phase structure of $\cN=2^*$ SYM on ellipsoids}

\author{Daniele Marmiroli}
\affiliation{Nordita, KTH Royal Institute of Technology and Stockholm University,\\Roslagstullsbacken 23 SE-106 91 Stockholm Sweden} 
\emailAdd{daniele.marmiroli@nordita.org}
\vspace{60mm}

\abstract{We analyse the phase structure of an $\cN=2$ massive deformation of $\cN=4$ SYM theory on an four-dimensional ellipsoid using recent results on supersymmetric localisation. Besides the 't~Hooft coupling $\gl$, the relevant parameters appearing in the theory and discriminating between the different phases are the hypermultiplet mass $M$ and the deformation (or squashing) parameter $Q$. The master field approximation of the matrix model associated to the analytically continued theory in the regime $Q\sim 2M$ and on the compact space, is exactly solvable and does not display any phase transition, similarly to $\cN=2$ $SU(N)$ SYM with $2N$ massive hypermultiplets. In the strong coupling limit, equivalent in our settings to the decompactification of the four-dimensional ellipsoid, we find evidence that the theory undergoes an infinite number of phase transitions starting at finite coupling and accumulating at $\lambda=\infty$. Quite interestingly, the threshold points at which transitions occur can be pushed towards the weak coupling region by letting $Q$ approach $2M$. \vfill}


\begin{document}

\texttt{\hfill NORDITA-2014-117}

\maketitle

\newpage

\section{Overview}

The holographic principle states a precise equivalence between string theory and gauge theory \cite{Maldacena:1997re,Gubser:1998bc,Witten:1998qj}, though only in a few cases a neat formulation of such correspondence is known. Namely, for gauge theories with maximal supersymmetry in their respective space-time dimension. On the other hand, the recent past bears witness of an increasing interest in extending our knowledge of holography to theories with a lesser amount of symmetries. In particular, a massive deformation of $\cN=4$ SYM theory in four dimensions was considered in \cite{Russo:2012kj,Buchel:2013id,Russo:2013qaa,Russo:2013kea,Russo:2013sba,Chen:2014vka} in relation with its dual, type IIB strings on the so called Pilch-Warner background \cite{Pilch:2000ue}. This theory, known as $\cN=2^*$ SYM, displays a very interesting and complicated phase structure in the large $N$ limit, characterised by fourth-order \cite{Russo:2013kea} phase transitions  starting at finite values of the 't~Hooft coupling constant and accumulating at infinite $\gl$ \cite{Russo:2013qaa}. These transitions are associated to the blowing up of nearly massless states that become dominant in the strong coupling phase and  reproduce, in a somewhat unexpected, way the beahviour predicted by the supergravity solution  \cite{Pilch:2000ue,Buchel:2000cn}.\\

Testing the holographic correspondence implicitly presumes some understanding of the gauge theory at strong coupling. Luckily enough, for theories with sufficient supersymmetry, localisation is a rather powerful tool
that allows exact and direct computations in field theory \cite{Witten:1988ze}. By exact we mean at any value of the coupling constant and by direct we mean to solve the  path integral without relying on any possible duality. Indeed, the partition function of $\cN=4$ SYM on $\mathbb{S}^4$ and that of its $\cN=2$ supersymmetric massive deformation have been localised to  matrix integrals in \cite{Pestun:2007rz}. Precisely on these results, the aforementioned investigation of the phase structure of $\cN=2^*$ SYM found its  foundations, allowing highly non-trivial checks of an underlying holographic principle for non-conformal theories. Far from being an isolated case, a similar structure of weak phase transitions accumulating at infinite coupling has been observed in three-dimensional massive Chern-Simons theories \cite{Barranco:2014tla,Anderson:2014hxa,Russo:2014bda}, five-dimensional SYM-CS \cite{Minahan:2014hwa} and in SQCD models \cite{Wadia:2012fr,Russo:2013kea,Barranco:2014tla}. \\

On the other hand, recent progresses in the formulation of gauge theories admitting rigid supersymmetry on curved spaces \cite{Festuccia:2011ws,Jia:2011hw,Samtleben:2012gy,Klare:2012gn,Dumitrescu:2012ha}  have allowed the authors of \cite{Hama:2012bg} to localise to  matrix integrals the partition functions of $\cN=2$ supersymmetric gauge theories on the four-dimensional ellipsoid. The aim of this paper is to investigate the structure of the massive $\cN=2^*$ SYM theory, defined in section \ref{N2star}, on such curved space, in particular in light of a possible new generalisation of the holographic duality. In order to do so we avail on standard methods for solving matrix models in the large $N$, master field approximation. Solving the problem then amounts to determining the density function $\rho(a_i)$ of the eigenvalues $a_i$ of the matrix field, after which the expectation value of any supersymmetric observables compatible with the localisation procedure can be determined by the classical average over the density $\rho(a_i)$ itself. For the conformal theory defined on the round four-sphere, the master field approximation of $\rho$ is well known. At weak coupling it behaves at the boundary of the eigenvalue support $\cC=[-\mu,\mu]$ according to an inverse square root law $\rho(a)\sim 1/\sqrt{\mu^2-a^2}$ and $\mu=\sqrt{\gl}/2\pi$. The most natural generalisation of this result is to understand the theory in the nearly-conformal and nearly-flat approximation. It is known that, as the hypermultiplet is given a small mass $M$, the maximum eigenvalue receives a contribution proportional to $M^2$ \cite{Russo:2013kea}. In section \ref{sec:weak coupling} we show that adding a small deformation of the background, that turns it into an ellipsoid with eccentricity $\sqrt{1-b^4}$, amounts to a redefinition of the mass $M^2\to M^2-(b-1)^2$ for $b\sim 1$.\\

At strong coupling this picture is substantially unaltered. The strong coupling large $N$ density of eigenvalues obeys the same equation of the Gaussian case \cite{Pestun:2007rz}, up to a rescaling of the coupling constant proportional to the square mass and deformation, similarly to \cite{Buchel:2013id}. The solution is therefore Wigner semi-circle law $\rho(a)\sim \sqrt{\mu^2-a^2}$  with $\mu = \frac{1}{2\pi}\sqrt{\gl\left(\frac{Q^2}{4}-M^2\right)}$, being $Q=b+b^{-1}$. This regime has also been investigated in \cite{Crossley:2014oea} in the $M=0$ case and we find perfect agreement with the results found there. At finite values of the coupling constant finding a solution for $\rho$ becomes a hard task, due to the number of parameters involved in the game and the complicated phase structure of the theory. Although we have not succeeded in finding an analytical solution for generic deformations, we can focus on certain corners of the parameter space where exact computations are viable. Under the squashing of the four-sphere, the vacuum expectation value of hypermultiplets masses gets shifted to  $m_{ij}^H=\left|a_i-a_j\pm \frac{Q}{2}\mp M \right|$, meaning that massless modes appear at the thresholds $2\mu= n\left|\frac{Q}{2} - M \right|$, with $n$ an arbitrary integer. Hence, fine tuning $M$ and $Q$, all the thresholds can be made arbitrary small, pushing phase transition close to $\gl=0$. In the limit where $Q\to 2M$, as we show in section \ref{sec:fine-tuning}, the theory is well described by the matrix integral studied in \cite{Kazakov:1998ji}, that in turn emerges from the large $N$ limit of a $0 + 1$ dimensional supersymmetric matrix quantum mechanics of $U(N)$ matrices. Quite interestingly, phase transitions disappear in this limit, and the theory can be solved exactly at arbitrary coupling. The absence of phase transitions in this regime is somewhat reminiscent of $\cN=2$ $SU(N)$ SYM theory on $\mathbb{S}^4$ with $2N$ massive hypermultiplets in the fundamental representation \cite{Russo:2013kea}, but we have not checked whether additional symmetries are recovered in these settings.\\

For general values of $M$ and $Q$ we provide numerical evidence that the phase structure of the theory on the deformed geometry mimics that of the one on flat space. Indeed, the pattern of repeating transitions is enhanced by the displacements of the effective masses by $\pm \frac{Q}{2}$ terms. We analyse this complicated structure in section \ref{sec:decompactification}. The phenomenon briefly outlined above has another interesting consequence. As it turns out, it is possible to push the critical values of the coupling constant towards the weakly coupled region by adjusting $Q$ and $M$, or otherwise stated, phase transitions can appear at fixed coupling in the flow from the flat geometry to the curved one. To our knowledge, this is a novel and distinguishing feature of the $\cN=2^*$ theory on the ellipsoid that can be relevant for determining a plausible holographic dual theory.


\section{The partition function of $\cN=2^*$ SYM on ellipsoids}
\label{N2star}

Availing on supersymmetric localisation, the authors of \cite{Hama:2012bg} have been able to compute the matrix model formulation of the partition function of SYM theories with at least $\cN=2$ supersymmetry on the ellipsoid defined by

\be
\frac{x_0^2}{R_0^2}+\frac{x_1^2+x_2^2}{R_1^2}+\frac{x_3^2+x_4^2}{R_2^2} = 1
\ee

As on flat space, the massive $\cN=2$ theory of our present interest is obtained by giving a mass term $M$ to the hypermultiplet of $\cN=4$ SYM and Yukawa couplings dictated by supersymmetry. We refer to this theory as $\cN=2^*$ SYM on the ellipsoid. In the strongly coupled regime this theory was, to some extent, the interest of \cite{Crossley:2014oea}. The partition function of the massive theory reads

\be
\label{partition-function}
Z^{\cN=2^*} = \int \dd \hat a_0\, \e^{-\frac{8\pi^2}{g^2}\Tr\hat a_0^2} Z^{\cN=2^*}_{\rm 1-loop} \left| Z^{\cN=2}_{\rm inst} \right|^2
\ee

being 

\be
\label{Z-one-loop}
Z^{\cN=2^*}_{\rm 1-loop} = \frac{ Z^{\rm vec}_{\rm 1-loop} }{Z^{\rm hyp}_{\rm 1-loop} } 
\ee

The geometric deformation acting on the round sphere $\mathbb{S}^4$ reflects into an algebraic deformation of the one-loop determinants by a term proportional to the squashing parameter $Q$. For the vector multiplet one has


\be
\begin{split}
Z_{\rm 1-loop}^{\rm vec} 
\equiv \prod_{\ga\in\Gd_+} \frac{1}{(\hat a_0 \cdot \ga)^2} \, \Y_b(\hat a_0 \cdot \ga)\Y_b(-\hat a_0 \cdot \ga)
\end{split}
\ee

where the product is restricted to the positive roots $\ga$ of the Cartan subalgebra $\Delta$ of the gauge group. For any massive hypermultiplet in some representation $\cR$ one has a factor of 


\be
\left[Z_{\rm 1-loop}^{\rm hyp} \right]^{-1} 
\equiv  \prod_{\rho\in\cR^+} \Y\left(\frac{Q}{2} + \hat M + \hat a_0 \cdot \rho \right)\,\Y \left( \frac{Q}{2} + \hat M - \hat a_0 \cdot \rho \right)
\ee

and again the product is restricted to positive roots $\rho$. The notation is as follows: $a_0$ is a Cartan subalgebra valued real matrix, $\hat a_0=\sqrt{R_1 R_2}\,a_0$, $M$ is the hypermultiplet mass, $\hat M=\sqrt{R_1 R_2}\,M$, $b=\sqrt{R_1/R_2}$ and $Q=b+\frac{1}{b}$ is the deformation parameter. The infinite products have been regularised using the $\Y_b$ function defined in appendix (\ref{Upsilon}) which differs from the one used in  \cite{Hama:2012bg} only by a marginal normalisation factor. In terms of the eigenvalues $\hat a_0\cdot \ga$ it is straightforward to write down the contribution of each adjoint vector multiplet and each adjoint or fundamental hyper multiplet

\be
\label{eq:multiplets-eigenvalues}
\begin{array}{ll}
{\rm adjoint~vector~multiplet} & \prod_{i<j} \frac{1}{(\hat a_i -\hat a_j)^2} \, \Y_b(\hat a_i -\hat a_j)\Y_b(-\hat a_i +\hat a_j)\\
{\rm adjoint~hyper~multiplet} & \prod_{i<j} \Y\left(\frac{Q}{2} + \hat M + \hat a_i -\hat a_j \right)\,\Y \left( \frac{Q}{2} + \hat M -\hat a_i +\hat a_j \right)\\
{\rm fundamental~hyper~multiplet} \quad & \prod_{i} \Y\left(\frac{Q}{2} + \hat M + \hat a_i \right)
\end{array}
\ee

Note in particular the presence of a factor in the term corresponding to the one-loop contribution of the vector multiplet that cancels the Vandermonde determinant. Lastly, notice that in (\ref{partition-function}) the instanton contribution $\left| Z^{\cN=2}_{\rm inst} \right|^2$ is given by Nekrasov partition function \cite{Nekrasov:2002qd} counting (anti-)self-dual instantons localised at the north(south) pole of the ellipsoid where the theory approaches the $\Omega-$deformed theory with $\gep_1=1/R_1$ and $\gep_2=1/R_2$. In the following discussion we will totally forget about the non-perturbative contributions to the partition function, which in the large $N$ limit has the form

\be
Z_{\rm inst}\sim \e^{-\frac{8\pi^2 kN}{\gl}}
\ee

being $k\in\mathbb{N}$ the instanton number. The classical instanton action can eventually be renormalised and similarly the instanton moduli space can blow up quicky enough to produce a finite instanton contribution to the total action. But as it was indeed discussed in \cite{Russo:2013kea}, instantons are always exponentially suppressed throughout the whole phase diagram of the $\cN=2^*$ theory on the round four-sphere, and we reasonably expect the same to hold here.  In appendix (\ref{sec:no-deformation-limit}) we argue that (\ref{eq:multiplets-eigenvalues}) are indeed the correct massive deformations of the massless $\cN=2$ theory on the ellipsoid and they give back $\cN=4$ on the round sphere when both the massive and the geometric deformations are removed.


\subsection{Low energy theory and the $\gb-$function}

From the direct analysis of the partition function along the lines of \cite{Pestun:2007rz}, one can harvest some information about the low energy dynamics of the theory. Because of the nature of the space-time there are various scales that can be chosen as natural energy scales: $R_0,R_1,R_2$ and $M$. The description of the theory at low energies in terms of a running coupling constant brakes down whenever the energy of interactions becomes comparable with the on-shell mass of hypermultiplets. Also, hard deformations of the four-sphere act as an infrared cut-off along certain directions, altering the dynamics at very low energies. Hence, and with a bit of tum'ah, we can say this description holds at energy scales $\Lambda$ such that $1/R\ll\Lambda\ll M$. When the deformation parameter takes its minimum, meaning $b=1$, the one-loop determinant factor reproduces the theory on the round $\mathbb{S}^4$, see appendix (\ref{sec:no-deformation-limit}).
It is clear that when the mass $M$ is much larger than 1, relations (\ref{doubGtoG}), (\ref{Upsilon}) and the asymptotic expansion (\ref{BdoubGasymp}) tell us that the coupling constant gets renormalised by a $\gb-$function similar to the one of  $N=2^*$ SYM on $\mathbb{S}^4$  and which shall actually asymptote the latter in the $b\to 1$ limit. To this end let's consider $\gb=\frac{R_1-R_2}{2R_2}$ so that for $R_1\gtrsim R_2$ we have $Q=2+\gb^2$. Then proceeding along the lines of (\ref{eq:inf-prod-no-def-limit}) and following, we have for the vector multiplet

\be
 Z_{\rm 1-loop}^{\rm vec} = \prod_{n> 0}\left[(n-1+ \ii\hat a_0\cdot \ga)(n+1+\gb^2 -\ii\hat a_0\cdot \ga)(n-1+\ii\hat a_0\cdot \ga)(n+1+\gb^2 -\ii\hat a_0\cdot \ga)  \right]^n
\ee

which at first non-trivial order in $\gb$ reads

\be
\begin{split}
&\prod_{n> 0} \left[(n-1+ \ii\hat a_0\cdot \ga)(n+1-\ii\hat a_0\cdot \ga)(n-1+\ii\hat a_0\cdot \ga)(n+1 -\ii\hat a_0\cdot \ga)\right]^n\\
& \times\left[ 1+ n\gb^2 \frac{2n+2}{(n+1)^2-(\ii\hat a_0\cdot \ga)^2}\right]
\end{split}
\ee

and then by definition of Barnes G-function

\be
 Z_{\rm 1-loop}^{\rm vec} = (1+2\gb^2) (\ii\hat a_0 \cdot \ga)^2 G(1+\ii\hat a_0 \cdot \ga)G(1-\ii\hat a_0 \cdot \ga)G(\ii\hat a_0 \cdot \ga+1)G(\ii\hat a_0 \cdot \ga-1)
\ee

where we have used the fact that the behaviour of the infinite product is essentially determined by the large $n$ terms, so that the equation above boils down to a rescaling of $Z_{\rm 1-loop}^{\rm vec}$ by $1+2\gb^2$. Analogously the contribution of the hyper multiplet gets rescaled by $\frac{1}{1+\gb^2}$. We are interested in the effect of geometric deformations to the running coupling, thus we need to expand for large hyper's masses $M\to\infty$ keeping the ratio $M/Q$ fixed. Note that in doing so, the latter might be big as well, meaning that we are not constraining $Q$ to acquire small values. Expanding $Z_{\rm 1-loop}^{\rm hyp}$ for large values of the argument through (\ref{BdoubGasymp}) and keeping term that are proportional to $(\ii\hat a_0\cdot \ga)^2$, we have the leading order asymptotic beahviour

\be
\begin{split}
& \log\left\{ \Y\left( \frac{Q}{2} + \ii M +\ii\hat a_0\cdot \ga \right)\,\Y\left( \frac{Q}{2} + \ii M -\ii\hat a_0\cdot \ga \right) \right\}\\
\if
=& -\log\Gg_2\left( \frac{Q}{2} + \ii M + \ii\hat a_0\cdot \ga\right) -\log\Gg_2\left( \frac{Q}{2} - \ii M - \ii\hat a_0\cdot \ga \right) \\
&-\log\Gg_2\left( \frac{Q}{2} + \ii M - \ii\hat a_0\cdot \ga \right) -\log\Gg_2\left( \frac{Q}{2} - \ii M + \ii\hat a_0\cdot \ga \right)\\
\fi
=& (\ii\hat a_0\cdot \ga)^2\,\log\left( \frac{Q^2}{4} +M^2 \right) - \left( \frac{Q^2}{4} + M^2 \right)\log\left( \frac{Q^2}{4} + M^2 \right) +  \cO(\ii\hat a_0\cdot \ga\,\log (\ii\hat a_0\cdot \ga))
\end{split}
\ee 

Opposedly to the $\mathbb{S}^4$ case, also the 1-loop contribution of the vector multiplets is deformed and needs to be expanded at large $Q$

\be
\label{vector-expanded}
\begin{split}
& \log\left\{ \Y\left( \ii\hat a_0\cdot \ga \right)\,\Y\left(\ii\hat a_0\cdot \ga \right) \right\}=\\
\if
= &-\log\Gg_2\left( \ii\hat a_0\cdot \ga \right) -\log\Gg_2\left(Q - \ii\hat a_0\cdot \ga \right) -\log\Gg_2\left(- \ii\hat a_0\cdot \ga \right) -\log\Gg_2\left( Q + \ii\hat a_0\cdot \ga \right)\\
\fi
=& -\log\left[\Gg_2\left(\ii\hat a_0\cdot \ga \right)\Gg_2\left(-\ii\hat a_0\cdot \ga \right)\right]+\half[(\ii\hat a_0\cdot \ga)^2 + Q^2] \log Q^2 +{\rm subleading} 
\end{split}
\ee 

There is still one piece to consider, namely $\Gg_2\left(\ii\hat a_0\cdot \ga \right)\Gg_2\left(-\ii\hat a_0\cdot \ga \right)$, as we can see from the last line of (\ref{vector-expanded}). Using the expansion of the double Gamma for small values of one of the parameters (\ref{doubGsmallparam}) we have

\be
\begin{split}
&\log\left[\Gg_2\left(\ii\hat a_0\cdot \ga \right)\Gg_2\left(-\ii\hat a_0\cdot \ga \right)\right] = 
 -\half \left(\ii\hat a_0\cdot \ga \right)^2 \log b^2 + \left(\ii\hat a_0\cdot \ga \right)^2 + \\
& + 2 \zeta_R'(-1) -\zeta_H'\left(-1, \frac{\ii\hat a_0\cdot \ga}{b} \right)-\zeta_H'\left(-1, -\frac{\ii\hat a_0\cdot \ga}{b} \right) + \log\Gamma\left(\frac{\ii\hat a_0\cdot \ga}{b}\right) + \log\Gamma\left(-\frac{\ii\hat a_0\cdot \ga}{b}\right)\\
 &  -\ov{12 b^2}\left( \psi\left(\frac{\ii\hat a_0\cdot \ga}{b}\right) +\psi\left(\frac{-\ii\hat a_0\cdot \ga}{b}\right) + 2\gamma_E \right)   +\log 2\pi + C_r
\end{split}
\ee

The remainder can be quickly estimated 

\be
C_r = \sum_{k=2}^\infty \frac{(-1)^{2k-1} B_{2k}}{2k(2k-1)}\left( \gz_H\left( 2k-1,\frac{\ii\hat a_0\cdot \ga }{b}\right) - \gz_R(2k-1) \right) \sim \frac{\ii\hat a_0\cdot \ga }{3240\,b}
\ee

as $b\to\infty$. This expression can be highly simplified. For large values of $b$ the Hurwitz and Riemann zeta functions cancel against each other, and the Gamma functions can be expanded as well leading to

\be
\log\left[\Gg_2\left(\ii\hat a_0\cdot \ga \right)\Gg_2\left(-\ii\hat a_0\cdot \ga \right)\right] = 
-\half \left(\ii\hat a_0\cdot \ga \right)^2 \log b^2 + \left(\ii\hat a_0\cdot \ga \right)^2 -2\log\left(\frac{\ii\hat a_0\cdot \ga}{b}\right)^2 + \cO(1)
\ee

All the terms which are quadratic in $a_0\cdot \ga$ contribute to the Gaussian integration, renormalising the coupling constant and giving rise to the running coupling. So, with respect to the undeformed case, for large values of the mass $M$ and the deformation $Q$ the $\gb-$function gets modified to

\be
\label{beta-function}
\frac{1}{\gl_R(\Gl)} = \frac{1}{\gl} - \frac{C_2}{8\pi^2}\log\frac{\frac{Q^2}{4} -M^2 R^2}{Q}
\ee

and now we have dimensionful quantities having restored powers of the radius $R$ (as a short notation for $\sqrt{R_1 R_2}$) and $C_2$ is the second Casimir of the gauge group. The energy scale $\Lambda$ is identified with the inverse size of the ellipsoid $1/\Lambda\sim\sqrt{R_1 R_2}$. Linear and logarithmic terms should not alter the convergence of the integral, and $a_0$ independent ones can be discarded in the overall divergent constant. Note that for the $N=4$ theory the  beta function boils down to a Q-dependent rescaling of the coupling constant.



\section{Almost conformal theory in the weakly coupled regime}
\label{sec:weak coupling}

There are several different regimes in which the theory can be understood analytically. To this end, it is often useful to separate the scales that appear in the game. Besides the hypermultiplet's mass $M$ which is set by hand into the theory, we must consider the width of the eigenvalue distribution $\mu$ , which is small at weak 't~Hooft coupling and increases at strong coupling, and the geometric deformation  parameter $b$. As anticipated, this gives rise to a rich ensemble of different behaviours. In section \ref{sec:decompactification} we derive a saddle point equation in the decompactification limit that takes into account all these effects simultaneously, though we are not able to solve such equation exactly, and we must rely on numerical results. In the present section we present analytical results obtained in different corners of the parameter space of the theory.

\subsection{Nearly conformal case}
\label{nearly-conformal-case}

In the following we mostly adopt conventions in which the Coulomb moduli $a_i$'s and the mass $M$ are dimensionless. As they appear ubiquitously in (\ref{partition-function}) and following equations, the moduli and masses are rescaled by the square root of the inverse product of the equivariant parameters $\gz= \sqrt{R_1\,R_2}$. However, in some circumstances it will turn out to be convenient to strip the hat off of such variables and make them dimensionless. This amounts to rescaling $Q$ as well

\be
\widehat a_i,\,\widehat M = \gz a_i,\,\gz M \rightarrow a_i, M; \qquad Q\rightarrow \widecheck Q=\frac{Q}{\gz} = \ov{R_1} + \ov{R_2}
\ee

with the effect that the nearly round case ($Q\gtrsim 2$) can be treated as a perturbation in $\widecheck Q$, assuming that $R_1,R_2 \bb 1$.
Exponentiating the one-loop partition function (\ref{Z-one-loop}) and then differentiating with respect to the Coulomb modulus one has the following saddle point equation for the matrix integral (\ref{partition-function})

\footnotesize
\be
\label{eq:saddle-weak}
\int_{-\mu}^\mu \dd \hat a_j\, \rho(\hat a_j)\,\bigg[ K(\hat a_i-\hat a_j)+K(-\hat a_i+\hat a_j) - K\left(\frac{Q}{2} + \hat M +\hat a_i-\hat a_j\right)
-K\left(\frac{Q}{2}+ \hat M -\hat a_i + \hat a_j\right) \bigg] = \frac{16\pi^2}{g_{YM}^2} \hat a_i
\ee
\normalsize

where $\mu$ is the width of the eigenvalues distribution and we have defined the Kernel function

\be
K(x) = \frac{d}{dx} \log\,\Y(x|b,b^{-1})
\ee

in (\ref{eq:kernel-definition}). At weak coupling the eigenvalue distribution approaches the Wigner semi-circle law in the same way as it does in the $\cN=2^*$ theory on the round sphere \cite{Buchel:2013id}

\be
\label{renormalised-mu}
\rho(x) = \frac{2}{\pi\mu^2}\sqrt{\mu^2-x^2} \quad {\rm with} \quad \mu = \frac{\sqrt{\gl}}{2\pi}\,f(Q,M)
\ee

hence the argument of the first and second contributions in (\ref{eq:saddle-weak}) above becomes small $\hat a_i-\hat a_j \sim \gz \mu\to 0$ as $\gl\to 0$, given that $\gz$ remains finite. We want to determine the function $f(Q,M)$, at least perturbatively in $Q$ and $M$ under the assumptions that $f(Q\sim 2,M\sim 0) = f_1(Q)+ M^2f_2(Q) $, and $f_2(Q)=$ a constant at lowest order. The additive structure can be inferred directly from the computations of \cite{Buchel:2013id}, whereas the factorisation of small $M$ corrections is implied by the fact that on the hard deformed ellipsoid the contribution of light fields must still be perturbative in the mass and nonperturbative in the geometry. To this end, we then use the asymptotic expansion of the double gamma function (\ref{eq:expansion-gamma-weak-coupling}), at first order in $a_i-a_j$ we can approximate

\be
\label{eq:massless-contr-to-saddle-point}
\begin{split}
 & K(\hat a_i-\hat a_j)+ K(-\hat a_i+\hat a_j) \\
=& -\frac{d}{d\hat a_i} \left[ \log \Gamma_2(\hat a_i-\hat a_j)+\log\Gamma_2(Q-\hat a_i+\hat a_j)+\log\Gamma_2(-\hat a_i+\hat a_j) + \log\Gamma_2(Q+\hat a_i-\hat a_j)\right]\\
\end{split}
\ee

with the leading contribution coming from the singularity in $a_i=a_j$

\be
\begin{split}
=& \frac{2}{\hat a_i-\hat a_j} -\frac{d}{d\hat a_i}\left( \log\Gamma_2(Q-\hat a_i+\hat a_j) + \log\Gamma_2(Q+\hat a_i-\hat a_j) \right)\\
=& \frac{2}{\hat a_i-\hat a_j} + \cO(\hat a_i-\hat a_j)
\end{split}
\ee

where the derivative contributions in the second to last line are at most $\cO(\hat a_i-\hat a_j)$ because of anti-symmetry. As expected, this gives rise to the undeformed saddle point equation only as long as $Q$ is not big enough for $\Gamma_2(Q)$ terms above to become non-negligible on the l.h.s. of (\ref{eq:saddle-weak}). In that case, and at first order in $\mu$,  they should be treated as constant that renormalises the r.h.s of (\ref{eq:saddle-weak}), along with akin contributions from $K(Q/2 \pm M)$. This is the same procedure that accounts for the running of the coupling constant in the $\cN=2$ theory on the four-sphere. 
Besides small mass corrections, we are interested in the contribution of small deformations of the $\mathbb{S}^4$. In the compact case, at zero mass $M$ and $Q=2$ the theory is conformal (\ref{beta-function}), therefore we can regard small $Q-2$ corrections as perturbations around the conformal theory. On the other hand, the general $Q$ case is hardly manageable, though interesting information can be extracted in certain peculiar limits, namely by means of fine tuning the hypermultiplets masses and the deformation parameter as in the upcoming section \ref{sec:fine-tuning}.


\subsection{Small deformations of the round geometry}
\label{sec:def-corrections}

The distribution of eigenvalues evidently deviates from the Wigner-Dyson law when we include subleading contributions at weak coupling. From the first and second summand in (\ref{eq:saddle-weak}) one can extract corrections proportional $a_i-a_j$ which in turn are given by higher powers of the argument in (\ref{eq:expansion-gamma-weak-coupling}). Note that $\cK$ in the equation cited above implicitly depends on $Q$ through (\ref{eq:kernel-definition}) even though its argument does not explicitly depend on it. The structure of corrections in this regime obeys

\be
\label{eq:first-subleading}
-\frac{d}{da_i} \left[ \log \Gamma_2(a_i-a_j)+\log\Gamma_2(-a_i+a_j) \right] =
 \frac{2}{a_i-a_j}+ \cA_k(a_i-a_j)^k
\ee

In order to determine the coefficients $\cA_k$ one has to show particular consideration to the regularisation of infinite products in $\Y$ functions. It turns out to be convenient to get rid of the divergent contributions to $\cK$ regularising the infinite products that appear in the $\Y$ function as in (\ref{eq:kernel-definition}), (\ref{derivative-on-gamma-exact}). Indeed, one is allowed to multiply $\Y(x)$ by a Gaussian factor to ensure convergence, without affecting the finite, $x$-dependent part. Setting 

\be
\label{sloppy-Psu-regularisation}
\Y(x) \to \prod_{mn} (mb+nb^{-1}+x)\e^{-\frac{\ga_{mn}x^2}{2}-\gb_{mn}x}\,\prod_{pq} (pb+qb^{-1}+Q-x)\e^{-\frac{\ga_{pq}(Q-x)^2}{2}-\gb_{pq}(Q-x)}
\ee

the extra $x$ dependent terms coming from the exponentials in $Z_{\rm1-loop}$ cancel between the vector and the hypermultiplet contributions. Moreover, accounting on (\ref{sloppy-Psu-regularisation}), the derivative in (\ref{derivative-on-gamma-exact}) gets modified to 

\be
\sum_{mn} (mb+nb^{-1}+x)^{-1} - x\,\sum_{mn}\ga_{mn}-\sum_{mn}\gb_{mn}
\ee

so that eventually the $\gz_R(1)$ divergences appearing in the $x\sim 0$ region  are cancelled by choosing $\ga_{mn}=\gb_{mn}^2=(mb+nb^{-1})^{-2}$. Interestingly, all the coefficients $\cA_k$ can be determined re-writing (\ref{eq:first-subleading}) through (\ref{derivative-logs-gamma}) as

\be
\begin{split}
&\sum_{m,n=0}^\infty \left[ \frac{1}{mb+nb^{-1}+a_{ij}}-\frac{1}{mb+nb^{-1}-a_{ij}}  \right] + 2a_{ij}\,\sum_{m+n>0} \frac{1}{(mb+nb^{-1})^2}\\
=&-2a_{ij}\, \left[ \sum_{m,n=0}^\infty \frac{1}{(mb+nb^{-1})^2-a_{ij}^2}- \sum_{m+n>0} \frac{1}{(mb+nb^{-1})^2}\right]\\
=& \frac{2}{a_{ij}} - 2a_{ij}\,\sum_{k=1}^\infty \left(\begin{array}{c}-1\\k\end{array} \right)(-1)^k a_{ij}^{2k} \sum_{m+n>0}\frac{1}{(mb+nb^{-1})^{2k+2}}
\end{split}
\ee

and since $\cA_1=0$ once the regularisation has been taken into account, they bring no linear correction to the saddle point equation. In order to compute corrections due to small deformations we set $b=1+\gb$ and $b^{-1}=1-\gb$ so that $Q=2+\gb^2$ and hence to second order in $\gb\ll1$ we see that all coefficients can be determined exactly as (\ref{sum-in-A1}) and following

\be
\label{all-A-coefficients}
\begin{split}
&\cA_{2k}= 0\\
&\cA_{2k+1} = 2(-1)^{k+1}  \left(\begin{array}{c}-1\\k\end{array} \right)\left\{ \gz_R(2k+1)+\gz_R(2k+2) - \gb^2   \left[(k+1) \left( \gz_R(2k+1)+\gz_R(2k+2) \right) \right. \right.\\
& \hspace{6ex}\left. \left. - \frac{(k+1)(2k+3)}{3}\left[\gz_R(2k) +4\gz_R(2k+1) +5\gz_R(2k+2)+2\gz_R(2k+3)  \right]  \right] \right\} 
\end{split}
\ee

for $k\geq 1$. The second subleading contribution to (\ref{eq:massless-contr-to-saddle-point}) can be computed the same way. Indeed, the small argument asymptotics of $\Gamma_2$ cannot be employed in the compact case since $Q$ is not, strictly speaking, a small quantity. Using  again (\ref{derivative-on-gamma-exact}) and proceeding as above, one finds for the unrenormalised coefficient

\be
\label{eq:second-subleading}
\begin{split}
\cB_k (a_i-a_j)^k
=&-2(a_i-a_j)  \sum_{m,n=0}^\infty \frac{1}{(mb+nb^{-1}+Q)^2}\left[1+  \frac{(a_i-a_j)^2}{(mb+nb^{-1}+Q)^2}  \right]\\
&+  \sum_{h>3}\cB_h (a_i-a_j)^h
\end{split}
\ee

Remarkably the leading contribution in $a_i-a_j$ can be determined nonperturbatively in $b$ in this case. Once the regularization  has been subtracted from (\ref{eq:second-subleading}) the divergent part is cancelled we are left with the finite contribution 

\be
\cB_1=- 2\left(b^2+b^{-2}\right)\gz_R(2)
\ee 

Higher orders in the weak coupling expansion can be easily computed in the $b\sim 1$ approximation. It is sufficient to note that they can be mapped to $\cA_k$ by identification of $m,n\to m+1,n+1$ and subtraction of the two terms with indices $\{m,n\}=\{0,1\},\,\{1,0\}$

\be
\label{A-B-relation}
\cA_{2k+1} + \cB_{2k+1} = 2(-1)^{k+1} \left(\begin{array}{c}-1\\k\end{array} \right)\left[ b^{2k+2}+b^{-2k-2} \right]\gz_R(2k+2)
\ee

Although the expression for $\cA_k$ given above has been obtained perturbatively in $\gb\sim 0$ and in the weak coupling regime, the relation  (\ref{A-B-relation}) is an identity to all orders in both $\gb$ and $a_i-a_j$. We can then re-sum the expansion and conclude that the first half of the kernel accounts for

\be
\label{half-kernel-re-summed}
K(\hat a_i-\hat a_j)+K(-\hat a_i+\hat a_j)=b\pi\,\cot((\hat a_i-\hat a_j)b\pi)+b^{-1}\pi\,\cot((\hat a_i-\hat a_j)b^{-1}\pi)
\ee  

where we have restored the dimensional dependence of the moduli. Now, since $b\gz=R_1$ and $b^{-1}\gz=R_2$, the latter equation suggests that the $\rho(a_i)$ is discontinuous at every point in the Coulomb moduli space where $|a_i|=\frac{n}{R_1},\,\frac{m}{R_2}$ with $n,m$ arbitrary integers. Although this equality cannot be fulfilled at weak coupling where the radia are of order unity and the eigenvalues are small, one can have a hint about the consequences of (\ref{half-kernel-re-summed}) in the decompactification limit. Namely as the $R$'s grow, also the size of the eigenvalue support grows, generating an infinite number of discontinuities. These appear as cusps at the ends of the support, moving towards the origin as $\gl$ is increased, and triggering the transition of $\rho$ from an inverse square root shape at weak coupling to a Wigner-like law at strong coupling. This is precisely the phenomenon described in \cite{Russo:2013qaa,Russo:2013kea} for massive $\cN=2$ theories on $\mathbb{S}^4$, the exception being that in the present case it emerges from geometrically altering the structure of space-time. It is then natural to expect that further adding a mass term for hypermultiplets will contribute with a second generation of discontinuities in the eigenvalue density at sufficiently large coupling.


\subsection{Conformal perturbations}

Next we must determine the $M$-dependent contribution to the saddle point equation coming from $K(Q/2 + M +a_i-a_j) + K(Q/2+M -a_i + a_j)$ in the kernel of (\ref{eq:saddle-weak}). There are again two different regimens in which analytic computations are viable. The first is the nearly conformal case of $M\sim 0$, the second is the large mass limit, which being equivalent to the decompactification limit will be treated separately. In the first case the mass $M$ is, for a compact ellipsoid, much smaller than $Q$ and comparable with $\mu$, so using the definition of $K$ and proceeding as in (\ref{eq:second-subleading}), one has

\be
\begin{split}
-\cC_k(a_i-a_j)^k =& K(Q/2 + M +a_i-a_j) + K(Q/2 + M -a_i + a_j) \\
=&-2(a_i-a_j+M)  \sum_{m,n=0}^\infty \frac{1}{\left( mb+nb^{-1}+\frac{Q}{2} \right)^2}\left[1+  \frac{(a_i-a_j+M)^2}{\left( mb+nb^{-1}+\frac{Q}{2} \right)^2}  \right] \\
&-2(a_i-a_j-M)  \sum_{m,n=0}^\infty \frac{1}{\left( mb+nb^{-1}+\frac{Q}{2} \right)^2}\left[1+  \frac{(a_i-a_j-M)^2}{\left( mb+nb^{-1}+\frac{Q}{2} \right)^2}  \right]\\
&-4(a_i-a_j)\sum_{m+n>0}^\infty \frac{1}{\left( mb+nb^{-1}\right)^2}  + \cO\left( |a_i-a_j|^3 +|M|^3 \right)
\end{split}
\ee

To the lowest non-trivial order in perturbation theory around $b\sim 1,\,M \sim 0$ it is easy to determine the behaviour of the regularised sums. First order terms in $M$ cancel so that to lowest non-trivial order the $M-$corrected contribution to the kernel is

\be
\begin{split}
\label{third-subleading}
& -\cC_1(a_i-a_j)\\
=& -4(a_i-a_j)  \sum_{\rm regularised} \frac{1}{\left( mb+nb^{-1}+\frac{Q}{2} \right)^2} - 12 M^2 \, (a_i-a_j) \sum_{m,n=0}^\infty \frac{1}{\left( mb+nb^{-1}+\frac{Q}{2} \right)^4}
\end{split}
\ee

where the regularised series is intended as subtracted of the last line of the equation above. Setting again $b=1+\gb$ and expanding around $\gb=0$, it is straightforward to determine the first few orders in the $\gb$ expansion (see appendix \ref{infinite-series}),

\be
\label{C1-perturbative}
\cC_1 = 4\gz_R(2) - 4\gb^2\left( 4\gz_R(2) + 3\gz_R(3) \right) +12 M^2 \left[\gz_R(3) + \half \gb^2 \bigg( \gz_R(3)-4\gz_R(5) \bigg) \right]
\ee

Quite interestingly, after the manipulations shown in (\ref{sum-in-C1}) in appendix, one can re-sum the series over the  indices $m,n$ and  rewrite it 

\be
\sum_{\rm regularised} \frac{1}{\left( mb+nb^{-1}+\frac{Q}{2} \right)^2} = \sum_{n=1}^\infty b^2 (-1)^{n+1} \psi^{(1)}(b^2 n+1) - \psi^{(1)}(2) 
\ee

being $\psi^{(n)}$ polygamma functions. We can now approximate this series with its dominant contribution. As $n$ grows, the $ \psi^{(1)}(b^2 n+1)$ is asymptotically $\frac{1}{b^2 n+1}$ and the series resums to the Hurwitz-Lerch Phi function. Eventually we get to 

\be
\label{C1-re-summed-in-beta}
 -4(a_i-a_j)  \sum_{\rm regularised} \frac{1}{\left( mb+nb^{-1}+\frac{Q}{2} \right)^2} = -4(a_i-a_j) \left[ \Phi(-1,1,b^{-2}+1) +1 - \gz_R(2) \right]
\ee

As we take the large deformation limit we see that this part of the kernel contributes with the asymptotic value

\be
 -4(a_i-a_j)  \left[ \log 2 +1 - \gz_R(2) -2\gb^{-2}\gz_R(2) + \cO\left( \gb^{-3} \right) \right]
\ee

therefore we conclude that for $b$ large enough and $M\sim 0$ the qualitative behaviour of the saddle point equation is entirely determined by (\ref{half-kernel-re-summed}). In principle one should be able to determine higher orders in $M$ and $Q$, though we see already from the expression above that the mass and the deformation parameter mix beyond the first non-trivial order. Moreover, none of these sums is expected  to receive contributions to linear order in a small $\gb$ expansion, due to the $b \leftrightarrow b^{-1}$ symmetry of the problem. We can now gather from (\ref{A-B-relation}) and (\ref{C1-perturbative}) all  $M^2$ and $\gb^2$ terms that are proportional to $a_i$ and hence appear on the r.h.s. of the saddle point equation and account for a rescaling of the coupling constant 

\be
\label{weak-coupling-rescaling}
\frac{8\pi^2}{\gl_R}a_i = \left[ \frac{8\pi^2}{\gl} + \left( \cA_1+\cB_1+\cC_1 \right) \right] \, a_i
\ee

from which 

\be
\label{f-of-Q-M}
f(Q,M) = 1+ \gl\frac{3\gz_R(3) \left[M^2-\gb^2 \right]}{8\pi^2}
\ee

In the latter equation the relative sign between $M^2$ and $\gb^2$ should not be suspicious, remember indeed that in our conventions $M$ is purely imaginary, while $\gb$ is real at this stage. 
Note also that for $b=1$ the latter reproduces, to this order, the result of \cite{Russo:2013kea}. Also, the structure of higher order corrections in (\ref{third-subleading}) agrees with what we have assumed at the beginning of this section based on general physical requirements on the gauge theory.

We conclude this section with a comment. In (\ref{half-kernel-re-summed}) we have pointed out that, whenever $R(a_i-a_j)$ equals an integer $n$ ($R$ is either of the radia), the leading contribution to the kernel of the saddle point equation is of the Hilbert kind. In addition note that the first correction is still linear in $a_i$, and reads

\be
\cA_1+\cB_1 = -2\gz_R(2)b^2
\ee

where we have assumed $b\bb 1$ for simplicity. In this regime the latter term dominates the r.h.s. of (\ref{weak-coupling-rescaling}) as the corrections due to (\ref{C1-re-summed-in-beta}) are bounded to approach a  constant. Henceforth

\be
\gl_R \sim \frac{8\pi^2}{b^2(2\gz_R(2))} \qquad {\rm for~} b\bb 1
\ee

can be interpreted as the statement that the theory squashed to the three dimensional sphere, and in the large radius limit, inevitably flows to the weakly coupled region. To make this point clearer, note that the resummation of perturbative contributions to first half of the kernel holds at finite values of  $a_i-a_j$, hence at finite coupling, while the contribution of the second half of the kernel are inversely suppressed as the $a_i-a_j$ grows. Moreover for $R\to \infty$ the condition $R(a_i-a_j)=n$ is fulfilled by an infinity of points in the Coulomb moduli space accumulating around $n=\infty$.  It follows that the dominant and first subleading behaviour of the rescaled coupling are entirely determined by the geometric collapse of the direction along the shrunk radius.


\section{Exact solution in the $M\to\frac{Q}{2}$ limit}
\label{sec:fine-tuning}

Although an analytic solution seems not to be viable for arbitrary hypermultiplets mass and deformation, one can think to extend one's comprehension of the dynamical properties of the theory beyond the small perturbation regime through a suitable {\it fine tuning} of $Q$ and $M$. It turns out to be interesting to analytically continue $M$ to real values and consider the  limit in which $M\sim Q/2$. Physical insight suggests that in this case the order of the $\gl\to 0$ and $M\to Q/2$ limits actually counts. More precisely, there must exist two different phases according to which of $\gl$ and $m=Q/2-M$ is smaller.  We first assume that both $\gl$ and $m$ are small compared to the other quantities entering the problem; then using the properties of Barnes double zeta functions, we can expand the kernel appearing in the saddle point equation according to

\be
\label{kernel-expansion-M-about-Q}
K \left( \frac{Q}{2} +M -a_i+a_j \right) = \frac{1}{a_i-a_j+\frac{Q}{2} - M }-\frac{d}{da_i}\log \Gamma_2\left( \frac{Q}{2} +M -a_i+a_j \right) + {\rm subleading}
\ee

and similarly for $K \left( \frac{Q}{2} +M +a_i-a_j \right)$.
The $\log\Gamma_2$ contributions that persist in these formulas are again worked out by means of (\ref{eq:second-subleading}), and since $M\sim Q/2$, exactly cancel the akin contribution coming from $K(a_i-a_j)+K(-a_i+a_j)$. It is straightforward to show that subleading terms also cancel and that the saddle point equation considerably simplifies to

\be
\label{eq:saddle-weak-M-about-Q}
\int_{-\mu}^\mu \dd a_j\, \rho(a_j)\,\bigg[\frac{1}{a_i-a_j}-\half\frac{1}{a_i-a_j+m}-\half \frac{1}{a_i-a_j-m} \bigg] = \frac{8\pi^2}{\gl} a_i
\ee

which holds up to corrections of order $a_i\,\left(\frac{Q}{2}+M\right)^{-1}$ to the right hand side and that can be reabsorbed in a redefinition of the coupling constant.
Quite interestingly the integral equation (\ref{eq:saddle-weak-M-about-Q}) emerges from the large $N$ approximation of the $0 + 1$ dimensional supersymmetric matrix quantum mechanics of $U(N)$ matrices interacting with some gauge field and its fermionic superpartner \cite{Kazakov:1998ji}. In turn, such theory is realised as the dimensional reduction of $\cN=1$ SYM from four to $0+1$ dimensions and describes the low energy dynamics of $N$ supersymmetric degrees of freedom probing a codimension-3 subspace of the full four-dimensional space. The solution to (\ref{eq:saddle-weak-M-about-Q}) for $m^2=-1$ has been determined analytically, in parametric form, in \cite{Kazakov:1998ji} developing a method originally proposed in \cite{Hoppe}. Since $Q$ is real and $M$ is purely imaginary we need to analytically continue $M$ to real values to define $m$. Physical values of the masses are then given by the analytic continuation of $m=\ii m'$ which though implies that $Q$ acquires unphysical values. The analytic continuation is perfectly well defined in the present case thanks to the analyticity properties of double gamma functions. In appendix (\ref{analytic-small-m}) we review the solution of \cite{Nekrasov:2002qd} in some details and add the dependence on $m$, emphasising certain aspects which are particularly relevant in our context. The solution of (\ref{eq:saddle-weak-M-about-Q}) is given by a set of parametric expressions for the rescaled 't~Hooft coupling $g^2=\frac{m'^2\gl}{8\pi^2}$, the maximum eigenvalue $\mu$ and the distribution $\rho(a_i)$.
Introducing by standard notation (\ref{elliptic-definition}) the incomplete/complete elliptical integrals of first and second kind $\mathbb{F}(\gamma,l),\,\mathbb{E}(\gamma,l),\,\mathbb{K}(l)=\mathbb{F}(\frac{\pi}{2},l)$ and $\mathbb{E}(l)=\mathbb{E}(\frac{\pi}{2},l)$, and the ratio $\gth(l)=\frac{\mathbb{E}(l)}{\mathbb{K}(l)}$, one finds (\ref{quantities-parametric-form}), (\ref{nu-1})

\be
\label{g-mu-and-nu-parametric}
\begin{split}
g^2(l) &=\frac{m'^2}{\pi^4}\chi^2(l) \mathbb{K}^4(l)\\
\mu(l) &=\frac{m'}{\pi} \left[ \mathbb{K}(l)\mathbb{E}(\gamma,l)-\mathbb{E}(l)\mathbb{F}(\gamma,l)   \right]\\
\nu^{(1)}(l) &= \frac{m'^2}{12}-\frac{2m'  \mathbb{K}^2(l) \theta (l)\big[ 5 \theta (l) \big( \theta (l)+l-2 \big)+(l-6) l+6)+(2-l) (l-1)\big]}{5 \pi ^2 \big[ \theta (l) \big( 3 \theta (l)+2 l-4 \big)-l+1 \big]}
\end{split}
\ee 

being the modulus $0<l<1$, the modular angle 

\be
\sin^2 \gg = \frac{\mathbb{K}(l)-\mathbb{E}(l)}{l\,\mathbb{K}(l)}
\ee

and the shorthand

\be
\chi^2(l)  = \frac{l(1-2\gth(l))-3\gth^2(l) +4\gth(l) -1}{3} 
\ee

Also, the momenta of the eigenvalue distribution are defined by $\nu^{(n)} = \int \dd x \rho(x) x^{2n} = \frac{1}{N} \left< \Tr\hat a_0^{2n} \right>$, and can be determined recursively up to (in principle) arbitrary order (\ref{momenta}). In particular $\nu^{(1)}$ is related to the derivative of the free energy with respect to the coupling

\be
\nu^{(1)} = \frac{g^4}{N^2}\frac{\d \mathcal{F}}{\d g^2}
\ee

The asymptotic analysis of (\ref{g-mu-and-nu-parametric}) is carried out in the following way. In the limit where $l\to 0$ the coupling goes also to zero, hence one can expand $g^2$ in powers of $l$, invert the series and substitute in the small $l$ expansion of the other relevant quantities. This produces a genuine weak coupling expansion. Analogously, the strong coupling asymptotics is computed by expanding $g^2$ around $l=1$. For the maximum eigenvalue this produces (\ref{mu-weak}) (\ref{mu-strong})

\be
\mu(\gl) = \left\{ \begin{array}{ll}
 m'\frac{\sqrt{\gl}}{2\pi}-\frac{m'}{\sqrt{2}} \left(\frac{\gl}{4\pi^2}\right)^\frac{3}{2} +\cO\left(\gl^5\right) & \gl\ll 1 \\
\frac{m'\sqrt[3]{3\pi^2\gl}}{4\pi} + \cO\left(\gl^{0}\right)
& \gl\bb 1  \end{array} \right.
\ee

Due to the nature of the correspondence with our original problem, we are interested in the weakly coupled expansion of $\mu$  in the range where the effective mass is bounded by $0<m<1$. By weak coupling here we mean small $g^2$, in such a way that $\mu=\sqrt{2}g+\cO(g^3)\ll 1$, so not necessarily small $\gl$. At weak coupling the density of eigenvalues behaves at the endpoints of the cut according to a square root law $\rho(x)\sim\sqrt{\mu^2-x^2}$ with $\mu^2$ increasing linearly as the 't~Hooft coupling and $m'^2$, which indeed plays the role of an effective mass term in an $\cN=2^*$-like theory on a round sphere. There are some interesting considerations one can draw from this simplified version of the original problem. Firstly, consider the $M\sim 0,\, b\sim 1$ expansion of section \ref{nearly-conformal-case} and the relative expression (\ref{f-of-Q-M}) for the function $f(Q,M)$ defined in (\ref{renormalised-mu}) as the first order deviation of the maximum eigenvalue from the round and massless background. Since the radius of convergence of the $M-$series is $M=1$ one can formally consider values of 

\be
M=\frac{Q}{2}-\tilde m \sim 1-\tilde m
\ee

with $\tilde m$ a now positive and small real number. Doing so one would get

\be
M^2-\gb^2 = 1 - 2\tilde m +\cO(m^2) 
\ee

and therefore $f(Q,M)$ would depend on one single parameter

\be
f(M= Q/2-\tilde m) = 1 - \gl\frac{3\gz(3)\,(1-\tilde m)}{4\pi^2}
\ee

in agreement with (\ref{mu-weak}), even though the precise expansions of $\mu(\gl)$ obtained in the two different ways above do not coincide due to the evident non-commutativity of the limits considered.  

Secondly, in the region where the 't~Hooft coupling is much smaller then the mass, the second and last terms under integral sign in (\ref{eq:saddle-weak-M-about-Q}) become subleading, due to the fact that $a_i-a_j\ll m'$, meaning that the equation simplifies further to

\be
\int_{-\mu}^\mu \dd a_j\, \rho_0(a_j)\,\frac{1}{a_i-a_j} = \frac{8\pi^2}{\gl_R} a_i \qquad \gl_R = \gl -\frac{\gl^2}{8\pi^2 m'^2}+\cO\left(\gl^3/m'^4\right)
\ee 

In this regime the exact solution is simply Wigner semicircle 

\be
\rho_0(x) = \frac{2}{\pi\mu^2}\sqrt{\mu^2-x^2}\qquad \mu=\frac{\sqrt{\gl_R}}{2\pi}  
\ee

Increasing the ratio of the 't~Hooft coupling against the effective mass one reaches a region in which the solution (\ref{g-mu-and-nu-parametric}) is the only good description. Further on, and since the kernel of the saddle point equation is only sensitive to the rescaled coupling, one can let $\gl$ take large values imposing the scaling limit $\gl\,m'^2\ll 1$. In these settings also $m'\ll\mu$ holds, meaning that the kernel above is approximately minus $m^2$ times the one-dimensional discrete Laplace operator acting on $\frac{1}{(a_i-a_j)}$

\be
\int_{-\mu}^\mu \dd a_j\, \rho_\infty(a_j)\,\frac{1}{(a_i-a_j)^3} = -\frac{8\pi^2}{m^2 g_{YM}^2} a_i
\ee

In this regime the maximum eigenvalue should be better described by (\ref{mu-strong}), but there exsist alternative descriptions of the master field solution, as shown in appendix \ref{more-solutions}. In particular one can notice that the analytic continuation from real to imaginary values of $m$ involves a phase transition from a Wigner-like distribution

\be
\rho_{\scriptsize R}(x) \sim \sqrt{\mu_{\scriptsize R}^2-x^2} \qquad \mu_{\scriptsize R}\sim \gl^{1/4}
\ee

 to solution with inverse square root behaviour at the boundary of the eigenvalue support

\be
\rho_{\scriptsize I}(x) \sim \frac{1}{\sqrt{\mu_{\scriptsize I}^2-x^2}}  \qquad \mu_{\scriptsize I}\sim \gl^{1/4}
\ee

Matrix models transitions of this kind are ususally well understood, and in the present case a better understanding of the phenomenon involved here can be relevant in uncovering the phase structure of the $\cN=2^*$ theory.
 
The analytic structure of the solution can be understood straightforwardly. As can be already seen from (\ref{g-mu-and-nu-parametric}), $\mu$ becomes purely imaginary when $m'$ is continued to give physical values of the deformation parameter. In other words, sending $m'\to 0$ the effective coupling constant $g^2$ goes to zero accordingly and the eigenvalue density gets squeezed in a region of zero size around the origin. Further continuing $m'$ along the imaginary axes produces negative values of $g^2$, though the 't~Hooft coupling $\gl$ remains positive. Moreover, the maximum eigenvalue $\mu$ becomes purely imaginary meaning that we entered an unphysical region. So, in these settings there is no phase transition at finite values of the coupling as the phenomenon of cut collision pointed out in \cite{Russo:2013qaa} happens at zero coupling and zero size.\\

Although the phase structure of the theory appears trivial in this {\it fine tuned} limit, there is some evidence that it is not. Let us consider the first momentum of $\rho(a_i)$ as given in (\ref{g-mu-and-nu-parametric}), its asymptotic expansions read at the first few orders

\be
\nu^{(1)}(\gl) = \left\{ \begin{array}{ll}
\frac{1}{12} \left(m'^2-m'^3\right)+\frac{\lambda  m'^3}{8 \pi ^2}-\frac{\lambda ^2 m'^3}{32 \pi ^4} +\cO\left(\gl^6\right)& \gl\ll 1\\
\frac{3}{5} \left(\frac{3}{\pi }\right)^{2/3} \lambda ^{2/3} m'^3+\frac{6 \sqrt[3]{3} \sqrt[3]{\lambda
   } m'^3}{5 \pi ^{4/3}}+\cO\left( \gl^0 \right) & \gl\bb 1  \end{array} \right.
\ee

\begin{figure}[htb]
\hrule
~\\
\begin{center}
\includegraphics[width=0.32\textwidth]{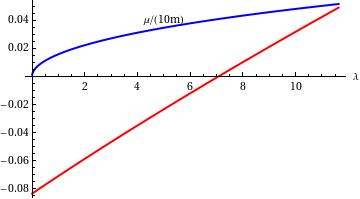}
\includegraphics[width=0.32\textwidth]{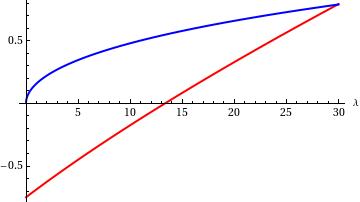}
\includegraphics[width=0.32\textwidth]{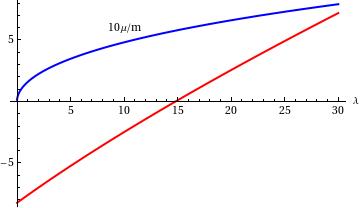}
\caption{\label{nu1-becoming-negative} In the analytically continued model with $m'\geq 1$ the expectation value of $x^2$ is not a positive semidefinite quantity for all values of the 't~Hooft coupling $\gl$. The red line represents $\left<x^2\right>$ as a function of $\gl$, while the blue line represents the rescaled maximum eigenvalue $\frac{\mu}{m'}$ for $m'=2,\,15,\,100$ respectively from left to right. The onset of a phase in which none of the scalar modes can be massless, otherwise stated $2\mu<m'$, is signaled by $\left<x^2\right>$ turning negative. Note that in the first (last) plot $\frac{\mu}{m'}$ is rescaled by $\frac{1}{10}$ (10).}
\end{center}
\hrule
\end{figure}

It is immediately clear from the weak coupling expansion that for $m'$ large enough one cannot consistently send $\gl\to 0$ as the semi-positive definite quantity $\nu^{(1)}$ apparently becomes negative, as can be seen in figure \ref{nu1-becoming-negative}. Interestingly there exists a physical interpretation of this fact, though, strictly speaking, it is not related to our original problem because of the fact that for $m'$ of order unity, the integral equation (\ref{eq:saddle-weak-M-about-Q}) ceases to be a good approximation of (\ref{eq:saddle-weak}). However, when $\gl$ is large enough  the width of the eigenvalue distribution is much larger than $m'$, meaning that the integral (\ref{eq:saddle-weak-M-about-Q}) receives most of its contribution from the singularities of the kernel due to modes of masses $a_i-a_j\pm n\,m',\,n\in \mathbb{N},$ that have become massless. If we gradually diminish $\gl$ we will eventually reach $2\mu=m'$, after which point the kernel is a regular function on the support of the integral. In figure \ref{nu1-becoming-negative} this phenomenon is evident and the second momentum (in red) becomes positive when the ratio of the maximum eigenvalue over the mass (in blue) approaches $\frac{\mu}{m'}=\frac{1}{2}$. 
Further decreasing $\gl$, in the region where $\gl\ll m'$ the Hilbert part of the kernel becomes largely dominant and the eigenvalues distribute according to the solution of the Gaussian model, Wigner semicircle law. This quite unexpected phenomenon is not evident for $m'<1$ \cite{Kazakov:1998ji} since the value at which $\nu^{(1)}$ turns negative is $g^2=1$ and hence outside of the domain of convergence of the weak coupling expansion.

\vspace{10mm}


\section{Decompactification limit}
\label{sec:decompactification}

\subsection{Asymptotic behaviour in the strongly coupled phase}
\label{asymp-strong-coupling}

Note that $\hat a_0 = \sqrt{R \tilde{R}}\, a_0= \frac{R}{b} a_0$ appears ubiquitously in the matrix model, therefore there are different scaling limits one can consider. One option is to keep $R$ fixed and send $\tilde{R}\to 0$; this produces $b\to\infty$ and accordingly $\hat a_0$ is always small. Henceforth in this particular scaling limit the matrix model can be always threated as it lays in the weakly coupled phase, assuming that the 't~Hooft coupling is bound to satisfy $\mu/b\ll 1$. We dub this phase the hardly deformed compact phase. A second option is to let $Q$ grow with one of the radia of the ellipsoid in a suitable decompactification limit by letting $R\to\infty$ and $\tilde{R}$ be finite. In this case $\mu$ grows with $\gl$. We refer to this as the hardly deformed decompactified phase, which in turn corresponds to the decompactification of the two dimensional theory obtained in the hardly deformed compact phase interacting with KK modes that propagate on the small compact circle of radius $\tilde{R}$. The third option is to consider the scaling limit $R/Q$ fixed, which also implies $\tilde{R}\sim1/R$, and that therefore describes the pure two dimensional theory on flat space. 

In the first case,  we can use the asymptotic expansion of Barnes double gamma from the very beginning and consider the scaling limit in which the difference between any two eigenvalues is in modulus much smaller then the deformation (rescaled by the radius)  $|\hat a_i-\hat a_j| \sim 2\hat\mu \ll Q$, accordingly 

\be
\begin{split}
& \frac{d}{d\hat a_i}\log \Gamma_2(Q-\hat a_{ij}) = \half Q\log Q -\hat a_{ij} \log Q -Q -\hat a_{ij} + \cO(\hat a^2)\\
& \frac{d}{d\hat a_i}\log \Gamma_2(Q+\hat a_{ij}) = -\half Q\log Q -\hat a_{ij} \log Q +Q -\hat a_{ij} + \cO(\hat a^2)\\
\end{split}
\ee 

For the mass dependent kernel functions we also have

\be
-K\left( \frac{Q}{2}+M+\hat a_{ij} \right) -K\left( \frac{Q}{2}+M-\hat a_{ij} \right)=-2\hat a_{ij} \log\left(\frac{Q^2}{4}-M^2  \right)
\ee

so that the all the information coming from squashing and mass terms is rephrased in a rescaled coupling constant on the r.h.s. of a saddle point equation of the form

\be
\int_{-\mu}^\mu \dd a_j\,\rho(a_j)\frac{1}{a_i-a_j} = a_i\left[\frac{8\pi^2}{\gl} +\log\frac{\frac{Q^2}{4}-M^2}{Q} \right]
\ee

Note that this integral equation for $\rho(a_i)$ is solved by Wigner semi-circle law, and that the beta-function such obtained coincides with the result found in (\ref{beta-function}) by direct inspection of the partition function.

On the other hand, in the second case pointed out above the contribution of $Q$ and $M$ to the kernel of (\ref{saddle-point-eq}) becomes subleading with respect to the Hilbert part, and in turn the saddle point equation can be approximated using (\ref{BdoubGasymp})

\be
\int_{-\mu}^\mu \dd \hat a_j\,\frac{\rho(\hat a_j)}{\hat a_i-\hat a_j}\left(\frac{Q^2}{4}-M^2 \right) = \hat a_i\,\frac{16\pi^2}{\gl} 
\ee

In \cite{Crossley:2014oea} the theory with $\cN=4$ supersymmetry was studied in this regime and the two descriptions agree when we send $M\to 0$ in the formula above.
Therefore the density of eigenvalues is again of Wigner type. So, in the two limiting cases above, equation (\ref{saddle-point-eq}) is solved by (\ref{renormalised-mu}) with

\be
\label{strong-coupling-mu-limits}
\mu(\gl,Q,M)= \left\{ \begin{array}{ll}
\frac{1}{2\pi} \sqrt{\frac{\gl}{1+\frac{\gl}{8\pi^2}\,\log\left(\frac{Q}{4}-\frac{M^2}{Q}  \right)}} & \qquad 2\mu \ll  {\rm min}(Q,M)\\
&\\ 
\frac{1}{2\pi}\sqrt{\gl\left(\frac{Q^2}{4}-M^2 \right)} & \qquad 2\mu \bb {\rm max}(Q,M)
\end{array}\right.
\ee

\begin{figure}[htb]
\hrule
~\\
\begin{center}
\includegraphics[width=0.45\textwidth]{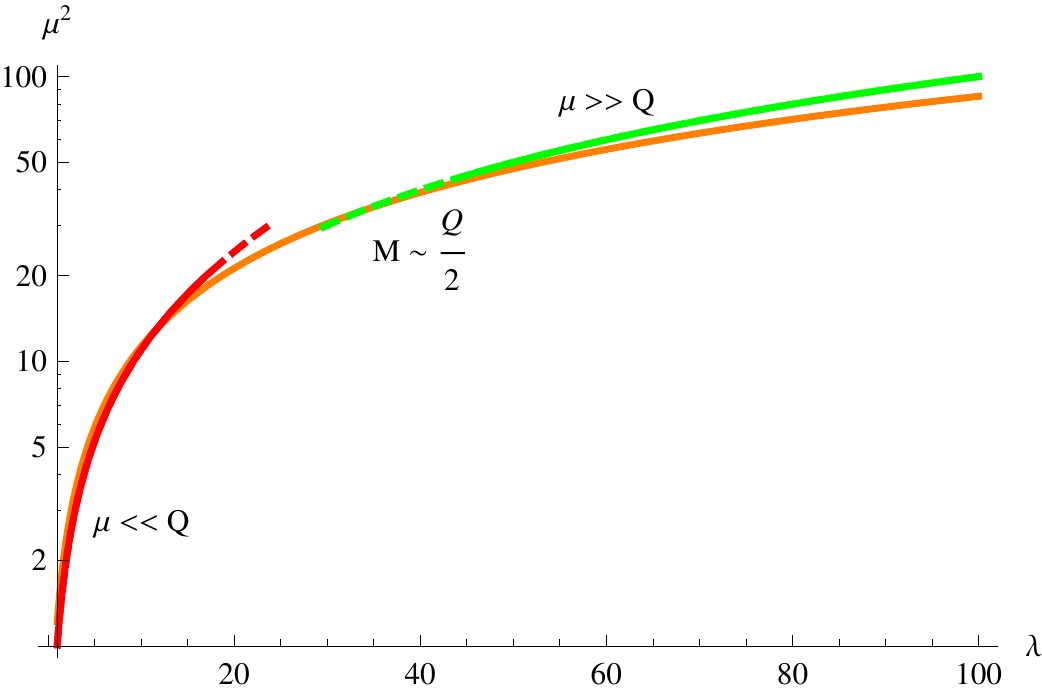}
\includegraphics[width=0.45\textwidth]{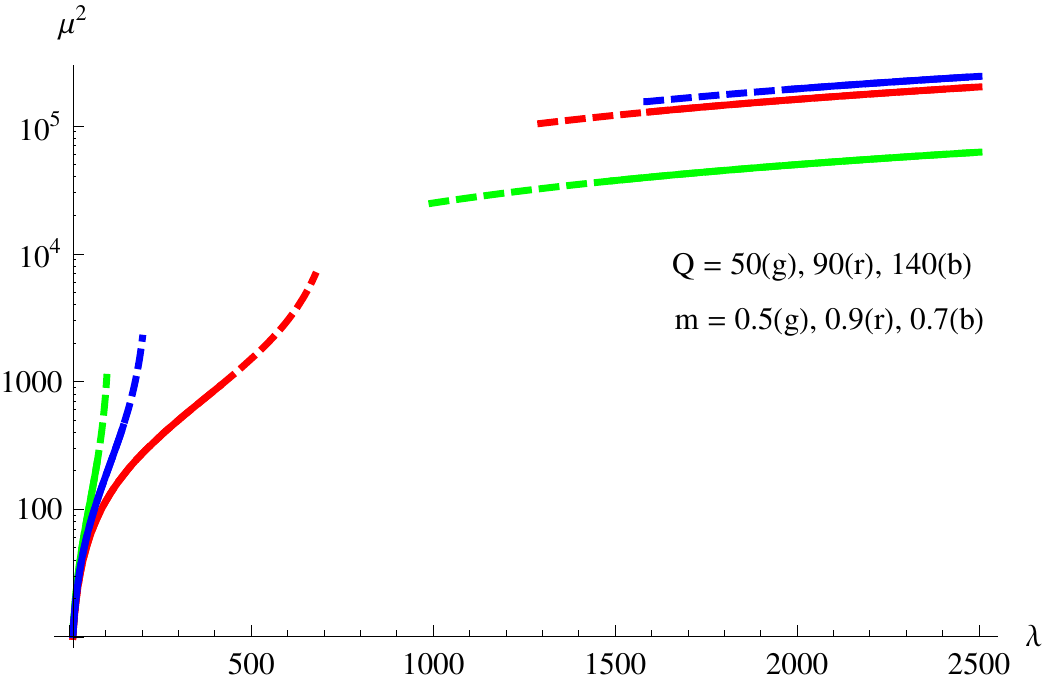}
\caption{\label{asymptotics-and-exact}  On the left: $\mu^2(\gl)$ (orange) as obtained solving exactly the fine tuned matrix model of section \ref{sec:fine-tuning} compared with the two asymptotic behaviours in the regions where  $\mu \ll Q$ (red) and $Q\ll\mu$ (green). The former matches the exact solution for $Q\sim 2M$, while the latter deviates significantly from it, due to large $Q$ corrections that make the integrand of (\ref{eq:saddle-weak-M-about-Q}) a bad approximation of the original problem. On the right: asymptotic expansions of the solution for increasing values of the effective mass $m=\frac{Q}{2}-M$. One can observe that as either $m$ or $Q$ are increased, the asymptotics deviates significantly from the simple behaviour depicted on the left. Note that ordinate axis are in logarithmic scale. }
\end{center}
\hrule
\end{figure}

The apparently implicit conditions on $\mu$ above can be worked out easily and set the ranges of $\gl$ in which the respective expansions hold. We can then consider the limit $m=\frac{Q}{2}-M\sim 0$ and compare the asymptotic solutions with that in (\ref{g-mu-and-nu-parametric}). As one can see from the left-hand plot depicted in figure \ref{asymptotics-and-exact}, the expansion at $\mu\ll Q$ (in red) is a good approximation of the exact solution (in orange) in the region where the maximum eigenvalue is itself smaller than one and the approximation of section \ref{sec:fine-tuning} holds. Quite interestingly the agreement between (\ref{g-mu-and-nu-parametric}) and the first of (\ref{strong-coupling-mu-limits})  is significant all the way through the weakly coupled region to the strongly coupled one, to some extent, thanks to the fact that the actual expansion parameter is $\mu=\frac{m\sqrt{\gl}}{2\pi}$, ($m=\half$ in the plot). For larger values of $m^2\gl$, the matrix model in (\ref{eq:saddle-weak-M-about-Q}) ceases to be a good description of the problem and the asymptotic solution at larger $\gl$ (in green) significantly differs from the exact solution of (\ref{eq:saddle-weak-M-about-Q}). In the right-hand plot of figure \ref{asymptotics-and-exact} one can observe the behaviour of the asymptotic solutions (\ref{strong-coupling-mu-limits}) at different values of $Q$ and $m$. In can be noticed that the bulk of the solution evidently gets stiffer by either increasing $Q$ or  $m$. This fact is symptomatic of the assumptions made in (\ref{eq:saddle-weak-M-about-Q}), in particular of the fact that such description breaks down when $Q-$dependent contributions resurges in the expansion (\ref{kernel-expansion-M-about-Q})  of the kernel of (\ref{eq:saddle-weak}) at very big values of $Q$. In this region there is no simple asymptotic description of the solution and one needs to study the complete saddle point equation in the decompactification limit, which is the subject of the next section.

\vspace{10mm}


\subsection{Strong coupling master field for general $Q$ and $M$}
\label{strong-coupling-general-Q-M}

In order to get some insight in the case of general $Q,\,M$ at large coupling we need to compute the logarithmic derivative appearing in the kernel of (\ref{eq:saddle-weak}). In order to do so, consider the definition of such functions

\be
\begin{split}
K(x) =& \frac{d}{dx} \log\,\Y(x|b,b^{-1})\\
 =&  \frac{d}{dx} \left\{ \log\Gg_2^{-1}(x|b,b^{-1}) + \log\Gg_2^{-1}(Q-x|b,b^{-1}) \right\}\\
=&  \frac{d}{dx} \left\{ -\gz_2'(0;b,b^{-1}|x) - \gz_2'(0;b,b^{-1}|Q-x) \right\}\\
\end{split}
\ee

where 

\be
\begin{split}
\frac{d}{ds}\gz_2(s;b,b^{-1}|x)\bigg|_{s=0} =& - \sum_{m,n} (mb+nb^{-1}+x)^{-s} \log(mb+nb^{-1}+x)\bigg|_{s=0}\\
=&-\log \prod_{m,n}(mb+nb^{-1}+x)
\end{split}
\ee

So deriving w.r.t. $x$

\be
-\frac{d}{dx}\gz_2'(0;b,b^{-1}|x) = \frac{\sum_{m,n}\prod_{i\neq n,j\neq n}(ib+jb^{-1}+x)}{ \prod_{m,n}(mb+nb^{-1}+x)}= \sum_{m,n}(mb+nb^{-1}+x)^{-1}
\ee

one arrives to the simple relation

\be
K(x) = \gz_2(1,b,b^{-1}|x)+\gz_2(1,b,b^{-1}|Q-x)
\ee

At this stage we need to reintroduce the dimensional dependence of $x$ on $R$ in order to extract information about the decompactification limit $R\to \infty$. After inverting the limit and the sum operations, one can approximate the double sum with a double integral over $\mu=m/R,\,\nu=n/R$ (note that this also rescales $K$ by $1/R$)

\be
\begin{split}
& \sum_{m,n\geq 0}^\infty \frac{1}{mb+nb^{-1}+xR} = \int_{1/R}^R\int_{1/R}^R \dd\mu\,\dd\nu\, \frac{1}{\mu b+\nu b^{-1} +x}\\
=& R\left[ Q\log Q+(b^{-1}-b)\log b \right] + x\log\frac{Q x}{R} -x +\cO\left(\frac{1}{R}\right)
\end{split}  
\ee

All of the four contributions to (\ref{eq:saddle-weak}) can be worked out this way. Putting all the terms together we see that pure divergences and linear terms globally cancel out. At the end of the day the contribution of $K$'s to the saddle point equation reads 

{\footnotesize
\be
\begin{split}
& K(Ra_{ij})+K(-Ra_{ij})-K(Q/2+RM+Ra_{ij})-K(Q/2+RM-Ra_{ij}) \\
=& \sum_{m,n}\left\{ \frac{1}{\Gw+Ra_{ij}}+\frac{1}{\Gw + Q-Ra_{ij}}+\frac{1}{\Gw-Ra_{ij}} +\frac{1}{\Gw+Q+Ra_{ij}} \right. \\
 &\left.-\frac{1}{\Gw+\frac{Q}{2}+MR+Ra_{ij}}-\frac{1}{\Gw+\frac{Q}{2}-MR-Ra_{ij}}-
\frac{1}{\Gw+\frac{Q}{2}+MR-Ra_{ij}}-\frac{1}{\Gw+\frac{Q}{2}-MR+Ra_{ij}}  \right\}
\end{split}
\ee
}

Though the latter gives rise to a particularly involved integral equation, let us point out that derivatives of $K(x)$ are much simpler

\be
K'(x)=\log\left(\frac{Q}{R}x\right)-\log\left(\frac{Q}{R}\left(\frac{Q}{R}-x\right)\right) \qquad K''(x)=\frac{1}{x}-\frac{1}{\frac{Q}{R}-x}
\ee

Hence we can differentiate twice the saddle point equation with respect to $x$ and get to

\be
\label{saddle-point-eq}
\begin{split}
&\fint_{-\mu}^\mu \dd y\, \rho(y)\left[ \frac{2}{x-y}+\frac{1}{x-y+\widehat Q}+\frac{1}{x-y-\widehat Q} \right.\\
& \left.-\frac{1}{x-y+\frac{\widehat Q}{2}+M} -\frac{1}{x-y+ \frac{\widehat Q}{2}-M}-\frac{1}{x-y-\frac{\widehat Q}{2} +M}-\frac{1}{x-y- \frac{\widehat Q}{2}-M}\right]=0
\end{split}
\ee

where $\widehat Q=Q/R$. Note that for large radius and $\widehat Q\sim 0$, this equation is equivalent to the undeformed case with mass $M$ studied in \cite{Buchel:2013id,Russo:2013qaa,Russo:2013kea,Chen:2014vka}. Indeed, this is the behaviour whenever the deformation scales slower then $R$ and is indeed killed by decompactification. On the other hand, also the opposite scaling limit, in which $Q$ is increased insanely faster then $R$, returns an equation of the same kind of that in \cite{Russo:2013qaa} with mass parameter $Q$. The solution in these two limiting cases is known exactly and one can draw some information from that, in particular we expect the solution to exhibit phase transitions in $\gl$ whenever a Coulomb modulus becomes null and a new massless boson appears. This corresponds to values of the coupling for which the width of the eigenvalue support hits an integer number times one of the mass shifts that appear in (\ref{saddle-point-eq})

\be
2\mu = nQ,\, n\left|\frac{Q}{2} \pm M\right| \qquad n \in \mathbb{N}
\ee 

and we immediately note that the first of such phase transitions can be arbitrarily close to $\gl=0$ as $2M$ approaches $Q$. At wak coupling the eigenvalue distribution has inverse square root singularities at the boundaries of the cut

\be
\rho_0(x) \sim \frac{1}{\sqrt{\mu^2-x^2}} \qquad \mu=\frac{\sqrt{\gl}}{2\pi}f(Q,M)  \qquad {\rm for~} |x|\sim \mu
\ee

and the maximum eigenvalues is proportional to $\sqrt{\gl}$ and to the function $f(Q,M)$ that we computed  in (\ref{f-of-Q-M}) for small values of the deformations. Introducing once more $m=\frac{Q}{2}-M$, we expect the first resonance to appear at $2\mu=m$ or, extrapolating from the weak coupling analysis, $\gl_c = m^2\pi^2$. As the coupling is further increased more resonances will appear and eventually the solution for $\rho(x)$ will approach Wigner semi-circle  distribution $\rho(x)\sim \sqrt{\mu^2-x^2}$ through the accumulation phenomenon pointed out in \cite{Russo:2013qaa}.


\subsection{$Q-$driven phase transitions}

The phase structure of the theory at $\widehat Q=0$ was understood in \cite{Buchel:2013id,Russo:2013qaa,Russo:2013kea,Chen:2014vka}. In this case there exists a sub-critical phase for values of the 't~Hooft coupling such that $2\mu(\gl)<M$ where the distribution of eigenvalues is dominated by an inverse square root law. For higher values of the coupling the maximum eigenvalue overcomes a new thresholds every time $2\mu=n M$, with $n\in\mathbb{N}$. At each threshold point a bosonic mode with Coulomb moduli $a_i-a_j= n M$ becomes massless and the theory enters a new phase. Eventually, for sufficiently large $\gl$, these contributions become dominant and the density of eigenvalues assumes the Wigner semi-circle shape typical of the type IIB supergravity solution.   Therefore, in order to understand how the deformation of the space-time alters this picture and what one can learn about a possible holographic dual candidate, we first consider small $\widehat Q$ perturbations around the known solution. Later we will let $\widehat Q$ become larger and eventually overcome $M$, which will allow us to gain a hint about the hardly deformed geometry.

\begin{figure}[h!]
\hrule
~\\
\begin{minipage}[l]{0.65\textwidth}
\includegraphics[width=.9\textwidth]{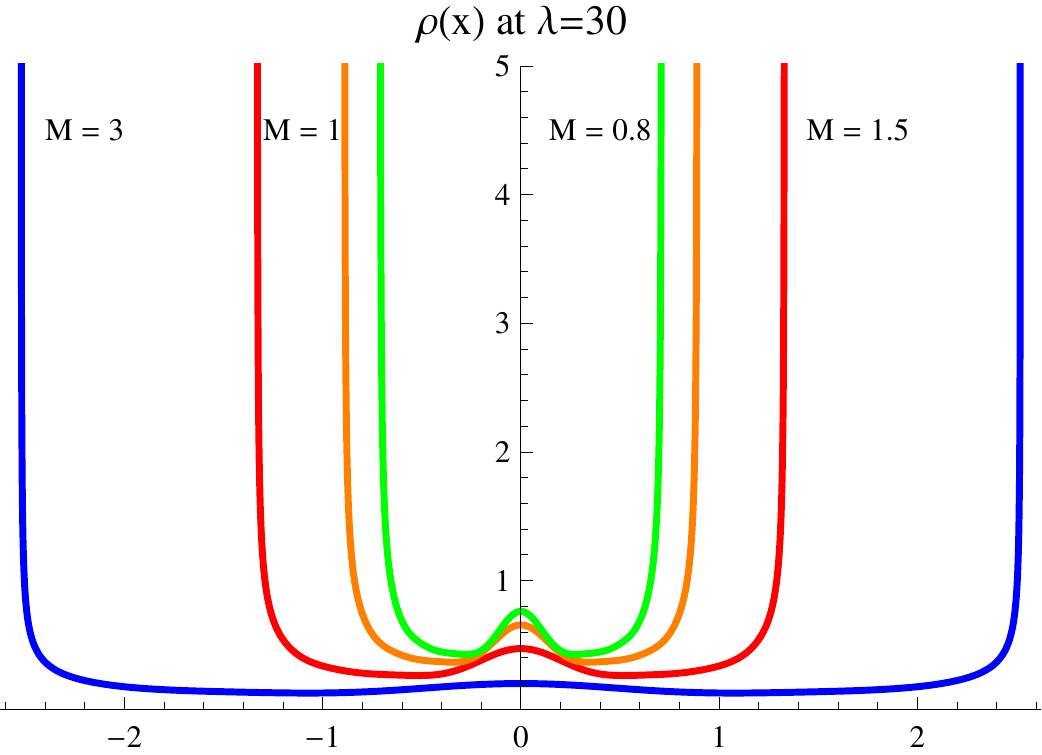}
\end{minipage}
\begin{minipage}[l]{0.35\textwidth}
\includegraphics[width=.9\textwidth]{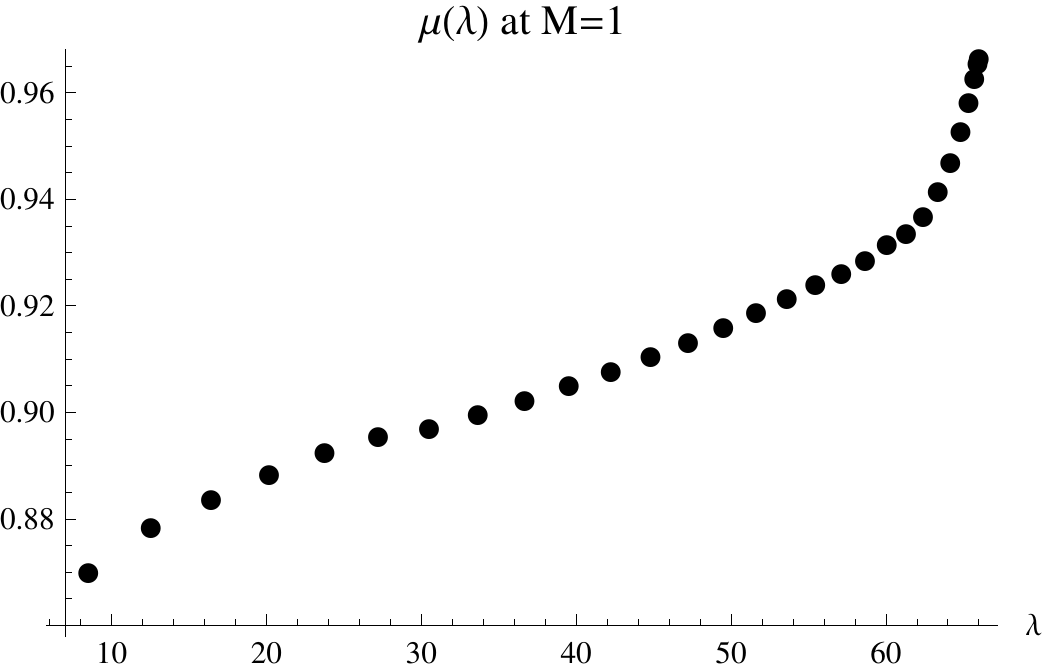}
\includegraphics[width=.9\textwidth]{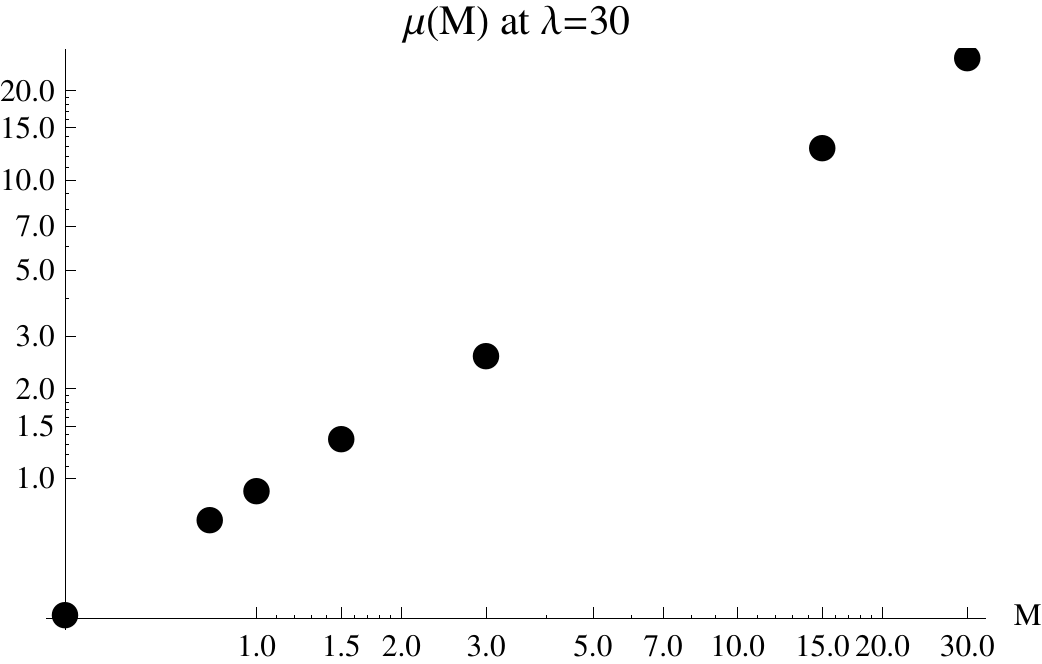}
\end{minipage}
\caption{\label{density-sub-critical-Q0} On the left: sub critical eigenvalue densities at different values of the hypermultiplet mass $M$ and negligible deformation $Q$ reproduce the known results obtained from the theory on the round $\mathbb{S}^4$. On the right: coupling dependence (on top) and mass dependence (bottom) of the maximum eigenvalue $\mu$. Dimensional analysis and asymptotic freedom imply that $\mu\sim M\e^{-8\pi^2/\gl}$ at weak coupling. The linear dependence on $M$ is evident in the lower graph. In the intermediate region the approximation done in (\ref{strong-coupling-mu-limits}) is in agreement with numerical results until $\gl \sim \frac{4\pi^2}{M}$. As the theory approaches the first threshold the maximum eigenvalue blows up and all the approximations done break down.}
\vspace{2ex}
\hrule
\end{figure}

\begin{figure}[h!]
\hrule
~\\
\begin{center}
\includegraphics[width=0.45\textwidth]{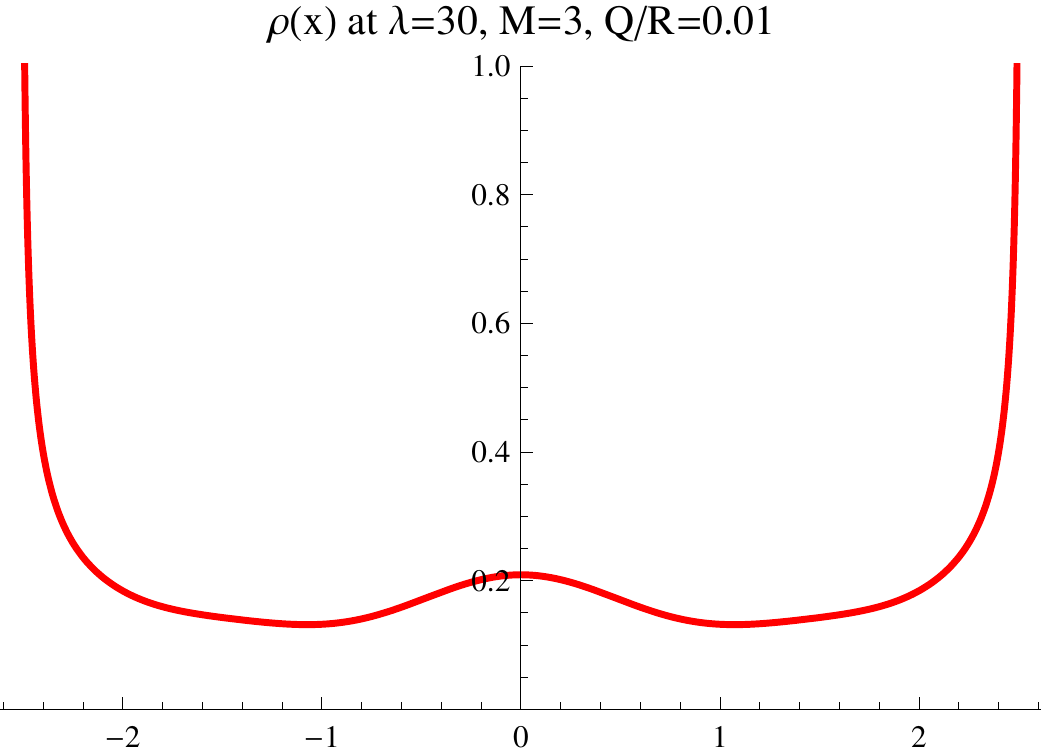}\hfill
\includegraphics[width=0.45\textwidth]{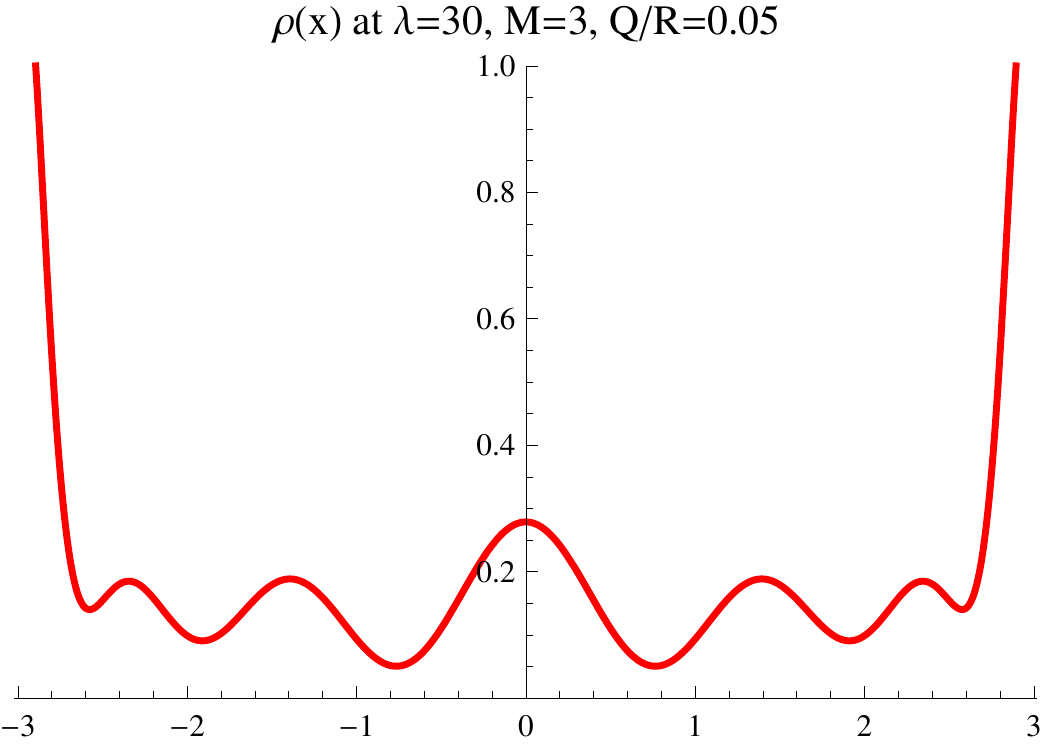}\vspace{5ex}
\includegraphics[width=0.45\textwidth]{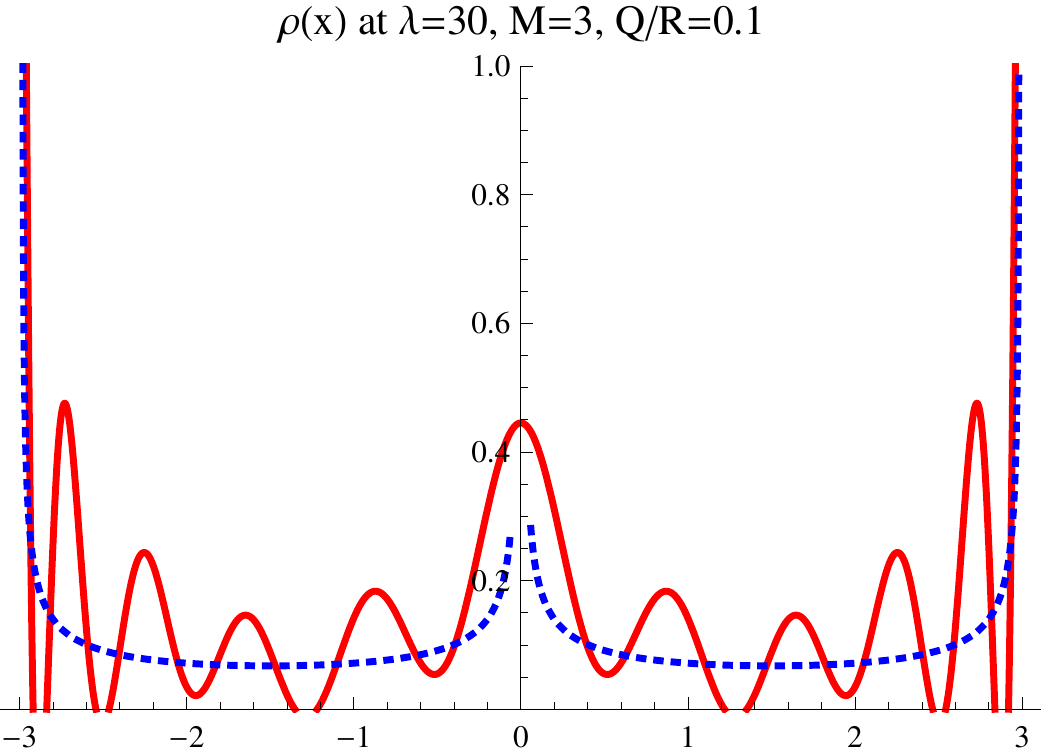}\hfill
\includegraphics[width=0.45\textwidth]{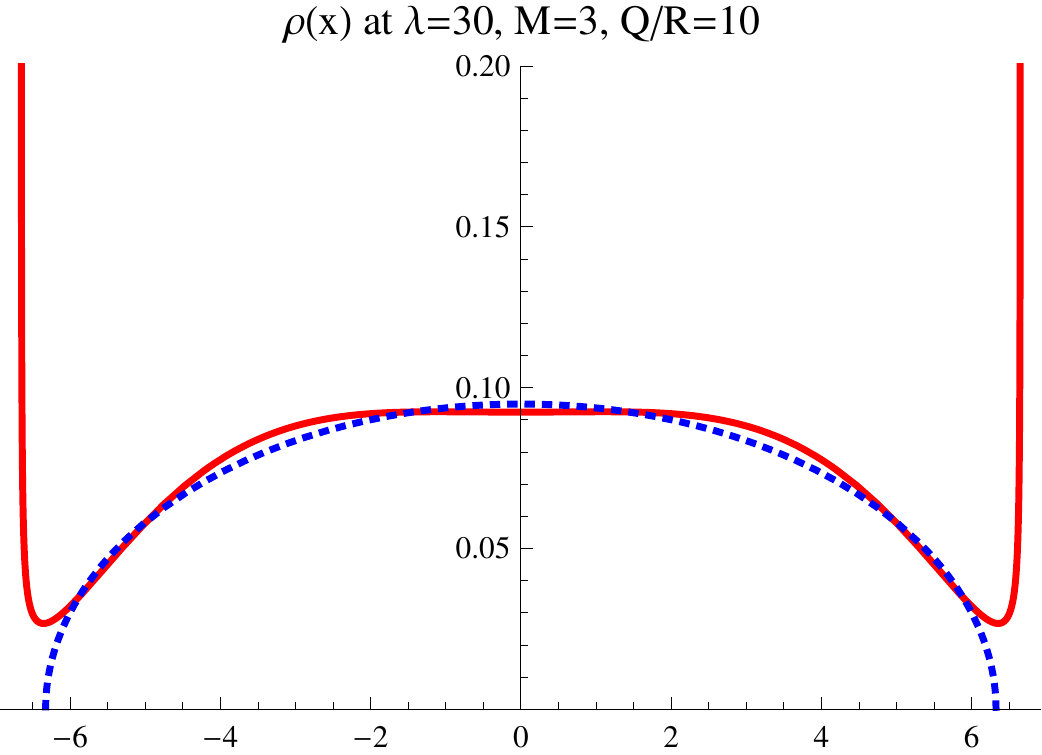}
\caption{\label{Q-perturbations-sub critical} Distribution of eigenvalues below the critical point, on the first line, and above the critical point, on the second line. Below the critical point the density of eigenvalues obeys an inverse square root law $\rho(x)\sim\sqrt{\mu^2-x^2}$. Criticality is achieved by increasing $Q$ in such a way that $\mu$ can overcome the first threshold located at $\mu=M-\widehat Q/2$. Below the critical point $Q-$oscillations are visible as superimposed to the inverse square root behaviour. Slightly above the critical point the first cusp-like transition appears as a coarse structure underling oscillations (dotted blue line). Further increasing $Q$, the maximum eigenvalue also increases, overcoming more thresholds and driving $\rho$ to a Wigner-like phase (dotted blue line).}
\end{center}
\hrule
\end{figure}

For $\widehat Q\sim 0$ the eigenvalue density is substantially determined by the ratio of $\mu/M$. Note that in this limit the coupling constant in (\ref{saddle-point-eq}) gets rescaled by a factor of two. One can then expect to find  a sub critical phase in which $\rho(x)$ still behaves as $1/\sqrt{\mu^2-x^2}$ at the boundary of integration.  Figures \ref{density-sub-critical-Q0} and \ref{Q-perturbations-sub critical} (top) show this is indeed the case, with a maximum eigenvalue 

\be
\mu\sim M f_\mu(\gl) \qquad f_\mu(\gl \ll M^2) \sim \exp\left[ -\frac{4\pi^2}{\gl}+\frac{3Q^2}{4M^4}\right]
\ee

which grows linearly with $M$ as dictated by dimensional analysis (fig. \ref{density-sub-critical-Q0} bottom-right). The function $f_\mu(\gl)$ is unknown for general values of $M$ and $Q$, though under the present approximations it is straightforward to see that it amounts to a rescaling of the coupling constant w.r.t. the $\widehat Q=0$. In figure \ref{density-sub-critical-Q0} (top-right) one can note the square root behaviour of $\mu$ for $\gl< \frac{4\pi^2}{M}$ ($\sim 40$ in the picture), in agreement with the approximation in (\ref{strong-coupling-mu-limits}), and its consequent blow-up as $\gl\to\gl_c$, being $\gl_c$ the critical value of the coupling at the first threshold.  \\

As the deformation is increased, $\mu$ grows quickly and the theory encounters a phase characterised by a highly oscillatory behaviour of $\rho$. In figures \ref{Q-perturbations-sub critical} top-right and bottom-left this behaviour is depicted, respectively, just below and above the critical point. In particular, one can notice that above the critical point, where $\mu$ has overcome $M-\frac{\widehat Q}{2}$, the oscillatory behaviour of $Q$ perturbations is superimposed to the cusp-triggered transition. The latter appears as a coarse structure (dotted blue line) underling smaller $Q-$oscillations (solid red line) and reproduce the structure of $\rho$ on flat space above the first threshold. The crucial point here is that the transition appears to be triggered by the increasing deformation and the coexistence of two phenomena: the increase of the maximum eigenvalue $\mu$ together with the decrease of the effective masses $|a_i-a_j|-|M-Q/2|$. This fact is a totally new feature w.r.t. the $\cN=2^*$ theory on flat space where transitions appear in the flow from the weakly coupled to the strongly coupled region. In addition, as $\widehat Q$ is increased further, the difference $M-\frac{\widehat Q}{2}$ becomes smaller with respect to $\mu$, meaning that an increasing number of massless bosons blow up. Eventually, these modes become dominant, letting the theory flow to a new phase with a Wigner-like distribution of eigenvalues characterised by 

\be
f_\mu(\gl \bb M^2) \equiv \sqrt{\gl}f_\mu(Q,M)=\sqrt{\gl}\left( 1- \frac{\widehat Q^2}{8M^2} \right)
\ee

as obtained in (\ref{strong-coupling-mu-limits}). This phase is strongly reminiscent of the strongly coupled phase of  $\cN=2^*$ theory on flat space. The phenomenon of cusp-accumulation that produces the transition is actually shared by the deformed and the un-deformed theories, though in the former, the effect of increasing the deformation reduces the distance between consecutive cusps at fixed coupling, while in the latter the coupling needs to be increased in order to reach the threshold. We can speak, in this case, of $Q-$driven phase transitions.\\

\vspace{10mm}

\begin{figure}[h!]
\hrule
~\\
\begin{center}
\includegraphics[width=0.45\textwidth]{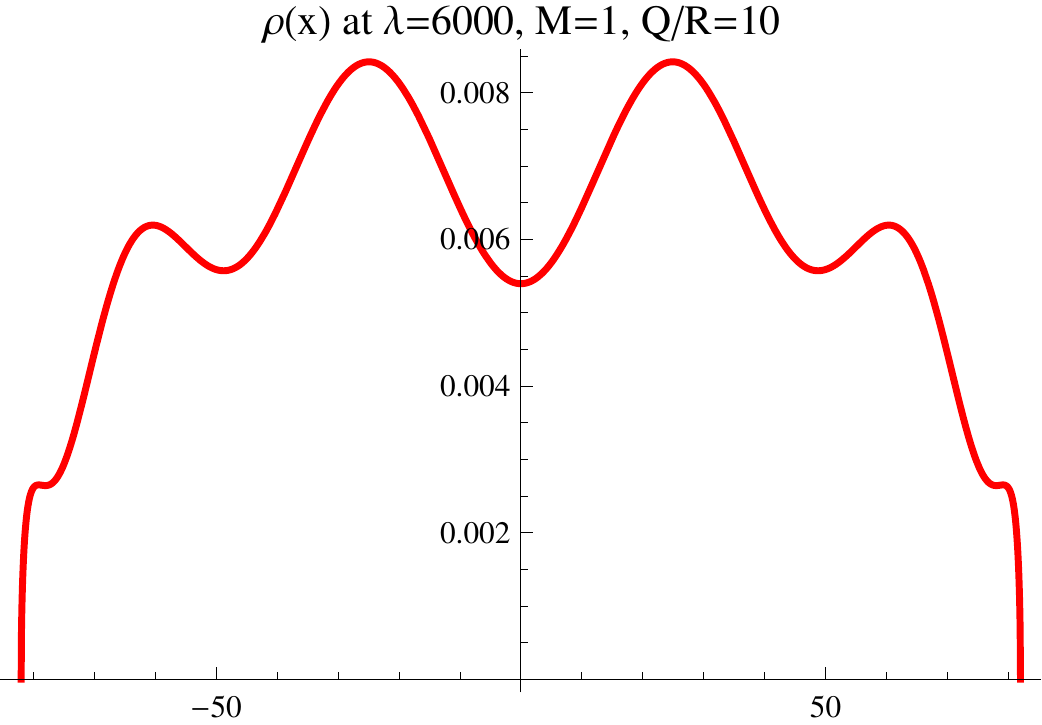}\hfill
\includegraphics[width=0.45\textwidth]{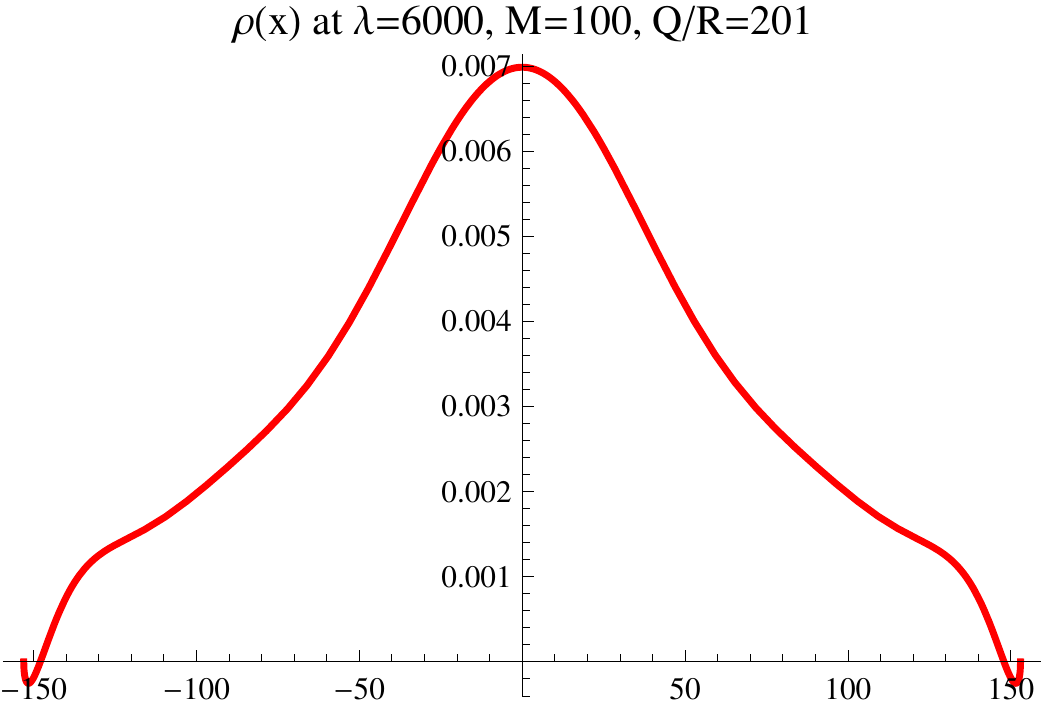}
\caption{\label{Q-supercritical}Distribution of eigenvalues far above the first critical point. On the left, for $\mu\sim 80$ and $\widehat Q/2-M=4$, cusp-driven phase transitions have brought the master field solution from the inverse square root behaviour to a Wigner-like distribution, deformed by $Q-$oscillations. On the right, small values of  $\widehat Q/2-M$ imply that the theory lies deeply in the strongly coupled region, while $Q$ comparable with $\mu$ cannot produce oscillations, resulting in just one peak on top of Wigner distribution.}
\end{center}
\hrule
\end{figure}

A way to have a deeper understanding of the $Q-$driven phase transitions  is to consider very large coupling $\gl$ and the limit $Q\to 2M$. In this regime the master field solution is asymptotically determined by (\ref{g-mu-and-nu-parametric}) and behaves at the boundary as $\sqrt{\mu^2-x^2}$. Figure \ref{Q-supercritical} depicts how, far above the critical point, the simple Wigner behaviour is altered by this phenomenon. For values of $Q,M\ll\mu$, on the left, the theory is dominated by massless modes and $\rho(x)$ approaches Wigner semicircle distribution, perturbed by heavy modes of large masses $M+Q/2$. As $Q$ is decreased further, the latter eventually decouple and such oscillations disappear. On the right of \ref{Q-supercritical},  values of $Q\sim 2M$, but still comparable with $\mu$ produce a big $Q-$oscillation, visible as a peak right in the middle of Wigners distribution. We can conclude that the ratio of $\mu/\widehat Q$ determines the entity of oscillations, while the ratio of $M/Q$ determines transition. These effects are particularly interesting in view of an holographic interpretation, even more so as they happen at fixed values of the coupling constant.

\section{Conclusions}

Evidently, $\cN=2^*$ SYM theory on an ellipsoid has an extremely rich structure. One can identify various regimes beyond the weakly and strongly coupled ones depending on the relative magnitude of the 't~Hooft coupling $\gl$, the squashing parameter $Q$ and the hypermultiplets mass $M$. Relying on supersymmetric localisation \cite{Hama:2012bg}, we have analysed in the details the contribution of small deformations to the weakly coupled, nearly conformal theory, and we have found agreement with the previously studied theory on the round $\mathbb{S}^4$ \cite{Russo:2013qaa,Russo:2013kea}. Also, we have identified a regime in which the dynamics of the non-conformal theory can be described in terms of $Q,M-$dependent $\gb-$function. One can immediately think of possible generalisations of these results to even less symmetric space-time manifolds. The next natural step would be to consider a squashed four-sphere, which in turn breaks an additional $U(1)$ symmetry. Indeed, Killing equations have been derived in \cite{Nosaka:2013cpa} but seem not to admit solutions in this specific case. \\

On the strong coupling side, though unable to find an analytic solution for arbitrary values of $Q$ and $M$, we are able to provide numerical evidence for a variety of new features that this theory exhibits. In some corner of the moduli space we are also able to provide an analytic understanding of these phenomena, and match previously established results. To our knowledge, a new and distinguishing feature of the analytically continued model is the presence of phase transitions at fixed coupling that are triggered by increasing the squashing $Q$ of the four-sphere.  These parallel the phase transitions at fixed $Q,M$ discovered in \cite{Russo:2013qaa} and that accumulate at $\gl=\infty$. The former seem to be particularly interesting in light of a possible holographic description, even more so as they can accumulate at finite values of the coupling constant when $M$ is fined tuned to be $\half Q/R$. At present, the meaning of such behaviour from the supergravity point of view is unclear, though it is clear that string fluctuations must be taken into account in order to probe such regime. \\

A number of non-trivial cross checks have been performed to match the $\cN=2^*$ SYM theory and its dual geometry \cite{Pilch:2000ue,Buchel:2000cn}, including the computation of large circular Wilson loops \cite{Buchel:2013id} and the free-energy of the theory on $\mathbb{S}^4$ \cite{Bobev:2013cja}. The strong coupling behaviour of large Wilson loops has also been reproduced in \cite{Bigazzi:2013xia}, where  a supergravity solution sourced by $D5-$branes wrapping an $\mathbb{S}^2$ has been considered. Although the squashed theory still lacks a string description, it is of primary interest, in order to shed some light on the nature of the latter, to consider similar observables at strong coupling from localisation, both in the compact and decompactified phases. Lastly, it should be possible to make contact between the theory in the compact phase under large deformations and two-dimensional Yang-Mills. This last aspect might be relevant in light of the known relations dictated by localisation between supersymmetric Wilson loops in four dimensional SYM and two-dimensional YM theory \cite{Drukker:2007yx,Pestun:2009nn}.

\section*{Acknowledgements}

I gratefully acknowledge The Foundation Blanceflor Boncompagni-Ludovisi, Nordita, KTH Royal Institute of Technology and Stockholm University for supporting my research. Also, I am thankful to Konstantin Zarembo for many useful discussions and comments about the draft of the present paper.


\appendix

\section{Barnes double zeta and related functions}
\label{sec:functions}

In this appendix we gather information about the special functions that appear throughout the text. 
Barnes double Zeta function is defined for any real number $x$, real positive $a,b$, natural numbers $m,n$ such that $ma+nb+x>0$ and complex $s$, as the infinite sum

\be
\label{BdoubZ}
\gz_2(x|a,b;s)= \sum_{m,n=0}^\infty \, (am+bn+x)^{-s}
\ee

The sum only converges for $\Re(s)>2$, but the function itself admits analytic continuations in all the parameters. A related function is Barnes double Gamma function

\be
\label{BdoubG}
\log\,\Gg_2(x|a,b)= \frac{d}{ds}\gz_2(x|a,b;s)\bigg|_{s=0} + \log\,\gr_2(a,b)
\ee

being 

\be
\log\,\gr_2(a,b)= -\lim_{x\to 0} \left( \frac{d}{ds}\gz_2(x|a,b;s)\bigg|_{s=0} +\log\,x \right)
\ee

Alternatively it is possible to show (Lerch formula)

\be
\log\,\Gg_2(x|a,b)= \frac{d}{ds}\left[\gz_2(x|a,b;s)-\chi(s;a,b)\right]_{s=0}  
\ee

where $\chi$ is doubled version of the Riemann zeta function

\be
\chi(s;a,b) = \sum_{m+n>0} (am+bn)^{-s}
\ee

The notation $m+n>0$ is intended as summing over all couples of non-negative integers that are not simultaneously null $(m,n)\in\cN^2_0-\{(0,0)\}$.
The function $\gz_2(x|a,b;s)$ has an analytic continuation in the complex $s-$plane with simple poles at $s=1$ and $s=2$. It follows that the Barnes double Gamma has the product representation 

\be
\label{BdoubGprod}
\Gg_2(x|a,b)= \frac{\e^{-r_{01}x + \half \left(r_{02}+r_{12}\right)x^2}}{x}\,\prod_{m+n>0}^\infty \,
\frac{1}{1+\frac{x}{\Gw}}\e^{\frac{x}{\Gw}-\frac{x^2}{2\Gw^2}}
\ee

where $\Gw=am+bn$ and $R_{ij}$ encode the residues ($i=1$) and finite parts ($i=0$) of $\gz_2(0|a,b;s)$ at the pole $s=j$. Their explicit expressions evade the present scope, so we avoid writing them. From the infinite product representation and the explicit form of the coefficients $R_{ij}$, it also follows the asymptotic expansion at large $x$

\begin{multline}
\label{BdoubGasymp} 
\log\,\Gg_2(x|a,b) =  -\frac{x^2}{2ab}\log x + \half\left( \frac{1}{a} + \frac{1}{b} \right)x\,\log x -  
\left[ \frac{1}{4} + \frac{1}{12} \left( \frac{a}{b} + \frac{b}{a} \right) \right]\log x\\
 + \frac{3x^2}{4ab} - \half\left( \frac{1}{a} + \frac{1}{b} \right)x - \chi'(0|a,b) + \cO(x^{-1})
\end{multline}

and at small $x$

\be
\label{eq:expansion-gamma-weak-coupling}
\log \Gamma_2(x|a,b) = -\log x - r_{01} x + \half\left( r_{02} + r_{12} \right) x^2 + \sum_{j=3}^\infty \frac{(-1)^j}{j} \chi(j;a,b)\, x^j
\ee

holding in the entire complex plane with a cut on the negative real axis.
The double Gamma can be expanded also for small values of its parameters (as well as for large values, but in our case there is no difference). For instance, for $a\to 0$ the following holds

\be
\label{doubGsmallparam}
\begin{split}
\log\,\Gamma_2(x|a,b)  \sim & \frac{b}{a}\left[\frac{x}{2b}\left(1-\frac{x}{b}\right)( \log\,b -1) - \gz_H'\left(-1,\frac{x}{b}\right) + \gz_R'(-1) \right]\\
&-\half\log\,a -\half \left(1-\frac{x}{b}\right) \log\,b + \half \log\,\Gamma \left(\frac{x}{b}\right) + \half\log 2\pi - \frac{a}{12b} \left(\psi\left(\frac{x}{b}\right) + \gamma_E \right)\\
&-\sum_{k=2}^\infty \frac{(-1)^{2k-1} B_{2k}}{2k(2k-1)}  \left(\frac{a}{b}\right)^{2k-1}  \left( \gz_H\left( 2k-1,\frac{x}{b}\right) - \gz_R(2k-1) \right)
\end{split}
\ee

Two more formulas are of particular interest to us. The first relates the double Gamma with unity parameters to the Barnes G-function

\be
\label{doubGtoG}
G(1+x)=\frac{\Gg(x)}{\Gg_2(x|1,1)}
\ee

while the second one is the shift formula

\be
\label{doubGshift}
\Gg_2(x+b|a,b) = \frac{ \sqrt{2\pi}\, a^{ \half-\frac{x}{a} } }{\Gg(\frac{x}{a})} \,\Gg_2(x|a,b)
\ee

We further introduce the $\Gg_b$ and $\Y_b$ function that appear in the one-loop regularized determinants of \cite{Hama:2012bg} and whose usage is widespread in the context of  Liouville CFT

\be
\begin{split}
\label{Upsilon}
\Y_b(x) &= \Gg_b^{-1}(x|,b,b^{-1}) \Gg_b^{-1}(Q-x|,b,b^{-1})\\
\Gg_b &= \frac{\Gg_2(x|b,b^{-1})}{\Gg_2(Q/2|b,b^{-1})}
\end{split}
\ee 

where $Q=b+b^{-1}$. From the meromorphic properties of the Barnes double Gamma it follows that the $\Y$ function is an entire function of $x$, has zeroes at $x=-\Gw,Q+\Gw$, and is symmetric under the exchange $b\to 1/b$. The $\Y$ function inherits its properties from the double Gamma and Zeta functions

\be
\label{Yproperties}
\begin{split}
& \Y_b(Q/2)=1 \\
& \Y_b(x) = \Y_b(Q-x) \\
& \Y_b(x+b)=\frac{\Gg(bx)b^{1-2bx} }{\Gg((1-x)b)}\,\Y_b(x) \\
& \Y_b(x+b^{-1})=\frac{\Gg(x/b)b^{2x/b-1} }{\Gg((1-x)/b)}\,\Y_b(x) 
\end{split}
\ee  

Also, it admits an integral representation in the strip $0<\Re(x)<Q$ 

\be
\Y_b(x)= \exp\int_0^\infty \frac{\dd t}{t}\,\left[ \left(\frac{Q}{2}-x\right)^2 \e^{-t}-\frac{\frac{t}{2}\,\sinh^2 \left(\frac{Q}{2}-x\right)}{ \sinh\frac{bt}{2}\,\sinh\frac{t}{2b}} \right]
\ee 

We refer the reader to the Appendix of \cite{Nakayama:2004vk} for a review of some of the properties of these special functions.


\section{Mapping infinite products}
\label{sec:no-deformation-limit}

In this section we provide evidence that the set of equations (\ref{eq:multiplets-eigenvalues}) represents the correct higgsing of the massless theory on the ellipsoid. We do so by showing that they reproduce $\cN=2$ theory on the ellipsoid, $\cN=2^*$ and $\cN=4$ on the round sphere when one removes respectively the mass deformation, the geometric one or both. Let us start from the contribution of the vector multiplet from \cite{Hama:2012bg} is

\be
\begin{split}
Z_{\rm 1-loop}^{\rm vec} = \prod_{\ga\in\Gd_+} \frac{1}{(\hat a_0 \cdot \ga)^2}
\prod_{m,n \geq 0}  & (mb+nb^{-1} + \hat a_0 \cdot \ga)(mb+nb^{-1} + Q- \hat a_0 \cdot \ga)\\
\times & (mb+nb^{-1} - \hat a_0 \cdot \ga)(mb+nb^{-1} + Q+ \hat a_0 \cdot \ga)
\end{split}
\ee

In their conventions $a_0$ is a Cartan subalgebra valued real matrix, $\hat a_0,\,\hat M$ habe been rescaled by $\sqrt{R_1 R_2}$, $b=\sqrt{R_1/R_2}$ and $Q=b+\frac{1}{b}$. The infinite product can be regularised using the $\Y_b$ function defined in (\ref{Upsilon})

\be
= \prod_{\ga\in\Gd_+} \frac{1}{(\hat a_0 \cdot \ga)^2} \, \Y_b(\hat a_0 \cdot \ga)\Y_b(-\hat a_0 \cdot \ga) 
\ee

Note that there is difference of $[\Gg_2(Q/2|b,b{-1})]^4$ in the normalisation between their conventions and the usual definition in CFT. It is of course unessential to keep track of it as it gets reabsorbed by a redefinition of the infinite multiplicative constant in front of the partition function. 

The limit $b\to 1$ corresponds to the theory on a round $\mathbb{S}^4$. In this limit the unrenormalised product reads

\be
\begin{split}
\label{eq:inf-prod-no-def-limit}
& \prod_{m,n \geq 0}  (mb+nb^{-1} + \hat a_0 \cdot \ga)(mb+nb^{-1} + Q- \hat a_0 \cdot \ga)\\
& \qquad \times(mb+nb^{-1} - \hat a_0 \cdot \ga)(mb+nb^{-1} + Q+ \hat a_0 \cdot \ga)\\
=& \prod_{n'\geq 0}  \left[(n' + \hat a_0 \cdot \ga)(n'+2 - \hat a_0 \cdot \ga) 
(n' - \hat a_0 \cdot \ga)(n'+2 + \hat a_0 \cdot \ga)\right]^{n'+1}\\ 
= & \prod_{n''> 0}  \left[(n''-1 + \hat a_0 \cdot \ga)(n''+1 - \hat a_0 \cdot \ga) 
(n''-1 - \hat a_0 \cdot \ga)(n''+1 + \hat a_0 \cdot \ga)\right]^{n''}\\ 
\end{split}
\ee

as there are $n'+1$ ways to write a non-negative integer $n'$ as the sum of two non-negative integers $m,n$. Rearranging the terms and using the $z\to-z$ symmetry of the $H(z)$ function, it is easy to convince oneself that this infinite product is indeed equivalent to

\be
(\hat a_0 \cdot \ga)^2 G(1+\hat a_0 \cdot \ga)G(1-\hat a_0 \cdot \ga)G(\hat a_0 \cdot \ga+1)G(\hat a_0 \cdot \ga-1) = (\hat a_0 \cdot \ga)^2\,H^2(\hat a_0 \cdot \ga)
\ee

One can work out the contribution of an hypermultiplet of mass $M$ in some representation $\cR$ in the same way

\be
\begin{split}
\left[Z_{\rm 1-loop}^{\rm hyp} \right]^{-1} =& \\
 = & \prod_{\rho\in\cR^+} 
\prod_{m,n \geq 0} \left(mb+nb^{-1} + \frac{Q}{2} + \hat M + \hat a_0 \cdot \rho \right) \left(mb+nb^{-1} + \frac{Q}{2} - \hat M - \hat a_0 \cdot \rho \right)\\
& \times \left(mb+nb^{-1} + \frac{Q}{2} + \hat M  - \hat a_0 \cdot \rho \right) \left(mb+nb^{-1} + \frac{Q}{2} - \hat M + \hat a_0 \cdot \rho \right)\\
= & \prod_{\rho\in\cR^+} \Y\left(\frac{Q}{2} + \hat M + \hat a_0 \cdot \rho \right)\,\Y \left( \frac{Q}{2} + \hat M - \hat a_0 \cdot \rho \right)
\end{split}
\ee

Note that, since $\Y(x)=\Y(Q-x)$, this is the same as

\be
\prod_{\rho\in\cR^+} \Y\left(\frac{Q}{2} + \hat M + \hat a_0 \cdot \rho \right)\,\Y \left( \frac{Q}{2} - \hat M + \hat a_0 \cdot \rho \right)
\ee

Since $a_0$ is real also $M$ is real. In the $b\to1$ limit the expression above simply reads

\be
\begin{split}
 & \prod_{\rho\in\cR} 
\prod_{n' \geq 0} \left[ \left(n'+1 + \hat M + \hat a_0 \cdot \rho \right) \left(n' + 1 - \hat M - \hat a_0 \cdot \rho \right) \right]^{n'+1}\\
& \times \left[ \left(n'+1 + \hat M  - \hat a_0 \cdot \rho \right) \left(n'+1 - \hat M + \hat a_0 \cdot \rho \right) \right]^{n'+1}\\
= & H\left(\hat M + \hat a_0 \cdot \rho \right)\, H\left(-\hat M + \hat a_0 \cdot \rho  \right)
\end{split}
\ee

Using the properties in (\ref{Yproperties}), it is possible to get rid of the $(\hat a_0\cdot\ga)^{-2}$ in the regularised one loop partition function and to obtain an expression

\be
\label{Z1loopshifted}
Z_{\rm 1-loop} = \frac{ \prod_{\ga\in\Gd_+} \Y(\hat a_0\cdot\ga+b)\, \Y(-\hat a_0\cdot\ga+b)\,b^{-2}}{\prod_{\rm hypers} \prod_{\rho\in\cR^+} \Y(Q/2 + \hat M + \hat a_0 \cdot \rho) \, \Y(Q/2 + \hat M - \hat a_0 \cdot \rho)}
\ee

for which the $b\to 1$ limit is straightforward. Then, availing on (\ref{BdoubGprod}), one can write the more compact expression

\be
\Y\left(x+\frac{Q}{2} \right) = \prod_{m,n} \left(\frac{\Gw'}{\Gw}\right)2 \left(1+\frac{x}{\Gw'}\right) \left(1- \frac{x}{\Gw'}\right) \, \e^{ \left(r_{01}-\frac{1}{\Gw}\right)Q- \left( r_{02}+r_{12} -\frac{1}{\Gw^2} \right) \left( x^2+\frac{Q^2}{4} \right) }
\ee

through which the overall, $a_0-$independent, multiplicative constant that makes (\ref{Z1loopshifted}) finite can be computed straightforwardly 

\be
\begin{split}
& \exp \left\{ \left( r_{02}+r_{12} -\sum_{m,n\geq 0} \frac{1}{\Gw^2} \right) \right. \\
 & \times \left. \left[ \left(\hat a_0\cdot\ga+b -\frac{Q}{2} \right)^2 +\left(-\hat a_0\cdot\ga+b-\frac{Q}{2} \right)^2 - \left(\hat M + \hat a_0 \cdot \rho \right)^2 - \left(\hat M - \hat a_0 \cdot \rho \right)^2 \right] \right\}\\
=& \exp\left\{ 2 \left( r_{02}+r_{12} -\sum_{m,n\geq 0} \frac{1}{\Gw^2} \right) \left( \left(b-\frac{Q}{2}\right)^2 -\hat M^2 \right) \right\}
\end{split}
\ee

where $\Gw'=\Gw+\frac{Q}{2}$ and we have assumed that the hypermultiplet comes in the same representation of the vector multiplet. Again we se that in the limit of no deformation $b\to 1$, Pestun's result is readily recovered.


\section{Kernel function and infinite series}
\label{infinite-series}

The kernel of the saddle point equation (\ref{eq:saddle-weak}) is conveniently defined through the function 

\be
\label{eq:kernel-definition}
\begin{split}
K(x) =& \frac{d}{dx} \log\,\Y(x|b,b^{-1})\\
 =&  \frac{d}{dx} \left\{ \log\Gg_2^{-1}(x|b,b^{-1}) + \log\Gg_2^{-1}(Q-x|b,b^{-1}) \right\}\\
=&  \gz_2(1;b,b^{-1}|x) + \gz_2(1;b,b^{-1}|Q-x) \\
\end{split}
\ee

the last equality holding by means of the definition of Barnes multiple gamma function

\be
\label{derivative-on-gamma-exact}
\begin{split}
&\frac{d}{dx} \log\Gamma_2(x|a,b) = \frac{d}{dx} \frac{d}{ds} \left[ \gz_2(x|a,b;s) - \chi(s;a,b) \right]_{s=0}\\
=& \frac{d}{ds} \frac{d}{dx} \sum_{m,n=0}^\infty (am+bn+x)^{-s} \bigg|_{s=0} = -\sum_{m,n=0}^\infty (am+bn+x)^{-1}\\
=&-\gz_2(x|a,b;1)
\end{split}
\ee

When inverting the order of the $x$ and $s$ derivatives, we always think about operating on the suitably regularised function, in which the $\gz_R(1)$ divergences appearing in the $x\sim 0$ region have been removed. See the discussion after (\ref{sloppy-Psu-regularisation}) about regularising these infinite sums. Using the equation above we can easily determine the small $x$ expansion of $K$

\be
\begin{split}
\label{derivative-logs-gamma}
&-\frac{d}{dx} \log\Gamma_2(Q-x)-\frac{d}{dx} \log\Gamma_2(Q+x)\\
=& -\gz_2(Q-x|b,b^{-1};-1) + \gz_2(Q+x|b,b^{-1};-1)\\
=&\sum_{m,n=0}^\infty \left\{ \frac{1}{mb+nb^{-1}+Q+x} -\frac{1}{mb+nb^{-1}+Q-x} \right\}\\
=&-2x  \sum_{m,n=0}^\infty \frac{1}{(mb+nb^{-1}+Q)^2}\left[1+  \frac{x^2}{(mb+nb^{-1}+Q)^2}  \right] + \cO\left( x^3 \right)
\end{split}
\ee

We can see from above that the coefficients of this expansions are typically infinite series involving two indices. In the general case, the sums of these series are rather hard to compute exactly, therefore we must exploit the dependence on the deformation parameter $b$ and expand  in a neighbourhood of $b$ in which they become treatable. The easiest of such series, once expanded around $b=1+\gb$ with $\gb\sim 0$, reads  

\be
\label{sum-in-A1}
\begin{split}
\sum_{m+n>0}^\infty \frac{1}{\left( mb+nb^{-1}\right)^s } \sim & \sum_{m+n>0}^\infty \left[ \frac{1}{\left( m+n\right)^s } - s\gb\frac{m-n}{\left( m+n\right)^{s+1} } \right] \\
&+ \gb^2 \sum_{m+n=0}^\infty \left[- s\frac{n}{\left( m+n\right)^{s+1}} + \frac{s(s+1)}{2}\frac{(m-n)^2}{\left( m+n\right)^{s+2}} \right]
\end{split}
\ee

The constant term is obtained simply by setting $k=m+n$ and counting multiplicities 

\be
\sum_{m+n>0}^\infty \frac{1}{\left( m+n\right)^s } = \sum_{k=1}^\infty \frac{k+1}{k^s} = \gz_R(s-1)+\gz_R(s) 
\ee

The first order contribution is zero because of anti-symmetry in $m\leftrightarrow n$, hence the first $\gb$-dependent correction is at order $\gb^2$, as suggested by the $b \leftrightarrow b^{-1}$ symmetry. The first sum at this order is easily seen to be, after symmetrisation, $-\frac{s}{2}$ times the zeroth order. The second sum is computed with the trick of determining the range of the difference $l$ of two given numbers whose sum $k$ is fixed, indeed $l$ takes the values $-k,\,-k+2,\,-k+4,\dots,\,k-2,\,k$, so

\be
\sum_{m+n>0}^\infty \frac{(m-n)^2}{(m+n)^{s+2}} = \sum_{k=1}^\infty \frac{1}{k^{s+2}}\sum_{l=-k/2}^{k/2}(2l)^2
=\frac{1}{3}\left[\gz_R(s-1) +\gz_R(s)+2\gz_R(s+1)  \right] 
\ee

Given the small $\gb$ expansion of (\ref{sum-in-A1}) it is in principle possible to compute all the contributions order by order in $\gb$, as the series appearing at higher order share this same structure. As one can se, there are divergences emerging at $s=2$, but they precisely cancel against the regularization terms. The second easiest sum is

\be
\label{sum-in-B1}
\begin{split}
\sum_{m,n=0}^\infty \frac{1}{\left( mb+nb^{-1} +Q \right)^s } \sim & \sum_{m,n=0}^\infty \left[ \frac{1}{\left( m+n +2 \right)^s } - s\gb\frac{m-n}{\left( m+n +2 \right)^{s+1} } \right] \\
&+ \gb^2 \sum_{m,n=0}^\infty \left[- s\frac{n+1}{\left( m+n +2 \right)^{s+1}} + \frac{s(s+1)}{2}\frac{(m-n)^2}{\left( m+n +2 \right)^{s+2}} \right]
\end{split}
\ee

and can be computing in the same way as above setting

\be
\sum_{m,n=0}^\infty \frac{1}{(m+n+2)^s} = \sum_{k=1}^\infty \frac{k}{(k+1)^s}=\sum_{k=1}^\infty \frac{k-1}{k^s}= \gz_R(s-1)-\gz_R(s)
\ee

The linear-in-$\gb$ term is again zero for parity reasons, and so are all odd-in-$\gb$ terms in the expansion. To compute terms proportional to even powers of $\gb$ we can use the same technique of (\ref{sum-in-A1}). Hence the first term appearing at order $\gb^2$ is again a half of $-\gz_R(s-1)+\gz_R(s)$. Among the infinite sums appearing above, the last one which is relevant for the case under study is

\be
\label{sum-in-C1}
\sum_{\rm regularised} \frac{1}{\left( mb+nb^{-1} +\frac{Q}{2} \right)^2 } = \sum_{m,n=0}^\infty \frac{1}{\left( mb+nb^{-1} +\frac{Q}{2} \right)^2 }-\sum_{m+n>0}^\infty \frac{1}{\left( mb+nb^{-1} \right)^2 }
\ee 

To this end it is convenient to employ the binomial expansion 

\be
\sum_{m,n=0}^\infty \frac{1}{\left( mb+nb^{-1} +\frac{Q}{2} \right)^2 } = \sum_{m,n=0}^\infty 	\sum_{k=0}^\infty 4b^2 \left( \begin{array}{c}-2\\k \end{array} \right) \frac{(b^2-1)^k (2m+1)^k}{(2m+2n+2)^{k+2}}
\ee

and then substitute $s=m+n$ as above and reorganise the series

\be
\sum_{s=0}^\infty \sum_{m=0}^s \sum_{k=0}^\infty \left( \begin{array}{c}-2\\k \end{array} \right)2^{-k}b^2 \frac{(b^2-1)^k (2m+1)^k}{(s+1)^{k+2}}
\ee

Note that the sum in $m$ is over all odd numbers. Proceeding as above for the contribution coming from the regularisation we obtain a sum over even numbers. Considering that the relative sign between these contribution is a minus, then we can write (\ref{sum-in-C1}) as

\be
(\ref{sum-in-C1}) = \sum_{s=0}^\infty \sum_{m=0}^{2s} \sum_{k=0}^\infty \left( \begin{array}{c}-2\\k \end{array} \right)2^{-k}b^2 \frac{(-1)^{m+1}(b^2-1)^k m^k}{s^{k+2}} =
\sum_{s=0}^\infty \sum_{m=0}^{2s} \frac{4b^2 (-1)^{m+1}}{(2s+(b^2-1)m)^2}
\ee 

At this stage it is clear how the divergent contribution is removed. To this end note that the double sum can be reorganised as

\be
\sum_{s=0}^\infty \sum_{m=0}^{2s} = \sum_{m=0}^\infty \sum_{s'=2}^\infty - \sum_{m=2}^\infty \sum_{s'=2}^{m}
\ee

where $s'=2s$, so that the region in the origin is removed. The series over $s'$ is (almost) the definition of polygamma functions

\be
(\ref{sum-in-C1}) =4b^2 \sum_{m=1}^\infty (-1)^{m+1} \psi^{(1)}(b^2 m+1)-4\psi^{(1)}(2)
\ee

The series above is dominated by its tail, and in the $m\bb 1$ region the argument behaves like

\be
\psi^{(1)}(b^2 m+1) \sim \frac{1}{b^2 m+1}
\ee

Therefore the last series can be rephrased using the definition of the Hurwitz-Lerch Phi function, and considering that $\psi^{(1)}(2)=\gz_R(2)-1$, eventually we can write

\be
(\ref{sum-in-C1}) = 4\left[\Phi(-1,1,b^{-2}+1) + 1 - \gz_R(2) \right]
\ee

Lastly, note that the for large $b$ the Phi function asymptotes a constant value $\Phi(-1,1,1)= \log 2$.


\section{Analytic solution of the compact model in the $M\sim\frac{Q}{2}$ limit}
\label{analytic-small-m}

The solution of the integral equation (\ref{eq:saddle-weak-M-about-Q}) was constructed in \cite{Kazakov:1998ji} availing on some older ideas of \cite{Hoppe}. We review such construction here and add the dependence on the parameter $m$. Later we analytically continue in the region where parameters acquire physical values and draw our conclusions. We will try to keep the notation as close as possible to the one in the original paper, and we refer the interested reader to the latter for all the details that we will skip here.

Let us first make a point about the notation. In our original problem the scalar fields take purely imaginary values, which thing is required by the convergence of the path integral, and in our notation they are expressed as $\ii \hat a_0$. Up to this point $a_0$ is a matrix in the Cartan subalgebra of the gauge group and the hat is the operation of multiplication by the square root of the inverse product of the two radia of the ellipsoid. It follows that the mass therm is also purely imaginary, but the deformation parameter $Q$ is by definition real and we will need to analytically continue over it. For the sake of neatness we hence choose the following notation - eigenvalues of the matrix $\ii \hat a_0$ are here denoted as $x,y$, the hypermultiplet mass $M$ is purely imaginary and $Q$ is real. The purpose of this section is to consider the particular limit in which $\frac{Q}{2}-M'=m\sim 0$, which makes proper sense only when we continue to $M'=\ii M \in \mathbb{R}$. In the limit we are considering the saddle point equation reads

\be
\label{saddle-again}
\int \frac{m^2\,\rho(y)\,\dd y}{(x-y)\left[ (x-y)^2-m^2 \right]}= -\frac{8\pi^2}{\gl} x
\ee

The problem where $m^2=-1$ was solved in \cite{Kazakov:1998ji}, therefore we set $m=\ii m'$ and consider real $m'$. To this end let us introduce the generalised resolvent

\be
\label{resolvent-definition}
\begin{split}
G(y) &= \frac{y^2}{g^2}+ \ii\left[ W\left(y+\frac{\ii m'}{2}\right)-W\left(y-\frac{\ii m'}{2}\right) \right]\\
W(z) &= \int_{-\mu}^\mu \dd y \frac{\rho(y)}{z-y}
\end{split}
\ee

which is related to the density of eigenvalues through

\be
\lim_{\gep\to 0} \left\{ W(y+\ii\gep)-W(y-\ii\gep) \right\}= -2\pi\ii \rho(y)
\ee

Considering then the following equation

\be
G\left(x+\frac{\ii m'}{2}\right)=G\left(x-\frac{\ii m'}{2}\right)
\ee

for $x$ belonging to the support of the eigenvalue distribution, by definition we get

\be
W(x+\ii\gep)+ W(x-\ii\gep)-W(x+\ii m') - W(x-\ii m') = \frac{2x}{g^2} \qquad {\rm for~}\gep\to 0
\ee 

that corresponds to (\ref{saddle-again}) with the identification $g^2=\frac{m'^2\gl}{4\pi^2}$. The definition of $G(z)$ implies it has two square root branch cuts over the intervals $z\in [ \pm\mu\pm\ii m'/2 ]$. Moreover one can convince oneself that the function $G(z)$ is real on the real $z-$axis, the imaginary axis and on the cuts. Thus, the problem of determining $\gz=G(z)$ is equivalent to the problem of finding the inverse map $\gz(z)$ that maps the upper half plane to the domain of reality of $G$. Since $G(z)$ is a holomorphic function on the upper-right quadrant deprived by the half cut $[\ii m'/2,\,\mu+\ii m'/2]$, the inverse map is uniquely determined by its Hilbert transform

\be
\label{inverse-map} 
z(\gz) = A \int_{x_1}^\gz \dd t\, \frac{(t-x_3)}{\sqrt{(t-x_1)(t-x_2)(x_4-t)}}
\ee 

where the turning points $x_1>x_2>x_3>x_4$ are the values of $G(z)$ respectively at $z=0,\,\ii m'/2 -\ii\gep,\, \mu+\ii m'/2, \,\ii m'/2+\ii\gep$. Imposing these actual values on $\gz(z)$ one is able to write a set of integral equations for the physical quantities

\be
\label{parameters-integral-form}
\begin{split}
\frac{m'}{2} &=  A \int_{x_2}^{x_1} \dd t\, \frac{(t-x_3)}{\sqrt{(t-x_1)(t-x_2)(x_4-t)}}\\
\mu &=  A \int_{x_3}^{x_2} \dd t\, \frac{(x_3-t)}{\sqrt{(t-x_1)(t-x_2)(x_4-t)}}\\
\mu &=  A \int_{x_4}^{x_3} \dd t\, \frac{(x_3-t)}{\sqrt{(t-x_1)(t-x_2)(x_4-t)}}\\
\end{split}
\ee

The remaining unknown quantities can be fixed by considering the large $\gz$ asymptotics (\ref{inverse-map}) and matching it with  large $z$ asymptotics of $G(z)$ in (\ref{resolvent-definition})

\be
\label{conditions-asymptotics}
\begin{split}
A &= \frac{g}{2}\\
\frac{g}{2} \int_{x_1}^\infty \dd t\,\frac{1}{\sqrt{t-x_1}} &= \gz(\infty)\\
x_1+x_2+x_4 &= 2x_3\\
x_1^2 + x_2^2 + x_4^2 -2x_3^2 &= \frac{6m'^2}{g^2}\\
m'^3-12m'\int x^2 \rho(x)\dd x &= 8 g^4 a
\end{split}
\ee

The last equation is particularly useful to determine the second moment of the eigenvalue distribution in parametric form in terms of the quantity 

\be
a = \frac{1}{40} \left(-x_1^3+\left(x_2+x_4\right) x_1^2+\left(x_2+x_4\right){}^2 x_1-\left(x_2-x_4\right){}^2 \left(x_2+x_4\right)\right)
\ee

The sets of equations (\ref{parameters-integral-form}) and (\ref{conditions-asymptotics}) fully solve the problem, albeit in a somewhat implicit way. To get some clue on the nature of the solution it turns out to be helpful to rephrase (\ref{parameters-integral-form}) in terms of elliptic integrals. Using the relations 

\be
\begin{split}
& \int_c^u \dd t\,\frac{t}{\sqrt{(a-t)(b-t)(t-c)}} = \frac{2a}{\sqrt{a-c}} \,\mathbb{F}(\gamma,l) - 2 \sqrt{a-c}\, \mathbb{E}(\gamma,l) \qquad a>b\geq u >c \\
& \int_c^u \dd t\,\frac{1}{\sqrt{(a-t)(b-t)(t-c)}} = \frac{2}{\sqrt{a-c}} \,\mathbb{F}(\gamma,l) \qquad a>b\geq u >c\\
& \int_u^a \dd t\,\frac{t}{\sqrt{(a-t)(t-b)(t-c)}} = \frac{2c}{\sqrt{a-c}}  \,\mathbb{F}(\delta,p) - 2 \sqrt{a-c}\, \mathbb{E}(\gd,p) \qquad a>u\geq b >c \\
& \int_u^a \dd t\,\frac{1}{\sqrt{(a-t)(t-b)(t-c)}} = \frac{2}{\sqrt{a-c}} \,\mathbb{F}(\gd,p) \qquad a>u\geq b >c\\
\end{split}
\ee

where 

\be
\label{elliptic-definition}
\mathbb{F}(\gamma,l) = \int_0^\gamma \frac{\dd \ga}{\sqrt{1-l\,\sin^2\ga}} \qquad \mathbb{E}(\gamma,l) = \int_0^\gamma \dd \ga\, \sqrt{1-l\,\sin^2\ga} 
\ee

are the incomplete elliptic integrals of, respectively, the first and second kind and 

\be
\gamma = {\rm arcsin}\sqrt{\frac{u-c}{b-c}} \qquad \gd = {\rm arcsin}\sqrt{\frac{a-u}{a-b}} \qquad l=\frac{b-c}{a-c}  \qquad l^2+p^2=1 
\ee

are their moduli, one obtains from (\ref{parameters-integral-form}) the set of parametric equations

\begin{gather}
\label{everything-elliptic-1}
(-x_1+x_2+x_4)\mathbb{K}(l) + 2(x_1-x_4)\mathbb{E}(l) = 0\\
\label{everything-elliptic-2}
(x_3-x_1)\mathbb{F}(\gamma,l) + (x_1-x_4)\mathbb{E}(\gamma,l) = 2\frac{\mu}{g}\sqrt{x_1-x_4}\,\\
\label{everything-elliptic-3}
2(x_4-x_3)\mathbb{K}(p) + 2 (x_1-x_4) \mathbb{E}(p) = \frac{m'}{g} \sqrt{x_1-x_4}
\end{gather}


In the first and last equation above we have used the fact that for $\gamma=\frac{\pi}{2}$ incomplete integrals become complete elliptic integrals $\mathbb{K}(l)= \mathbb{F}(\pi/2,\,l)$ and $\mathbb{E}(l)=\mathbb{E}(\pi/2,\,l)$. It is convenient to introduce new variables 

\be
\gl_i=\frac{x_i}{x_1-x_4} \qquad y_i=g\,x_i
\ee

for which one can easily see that 

\be
l=\gl_2-\gl_4 \qquad 1-l = \gl_1-\gl_2
\ee

The last $\gl_i$ is fixed by the third equation in (\ref{conditions-asymptotics}), while equation (\ref{everything-elliptic-1}) determines parametrically

\be
\gl_2=1-2\gth(l)=1-2\frac{\mathbb{E}(l)}{\mathbb{K}(l)}
\ee

Using these relations it is also straightforward to express the modular angle in (\ref{everything-elliptic-2}) in parametric form

\be
\sin^2 \gg = \frac{\mathbb{K}(l)-\mathbb{E}(l)}{l\,\mathbb{K}(l)}
\ee

At this stage all quantities are fixed in terms of one single parameter, the elliptic modulus $l\in[0,1]$, and it is just a matter of algebra to re-write (\ref{everything-elliptic-2}) and (\ref{everything-elliptic-3})  as

\be
\label{quantities-parametric-form}
\begin{split}
g^2(l) &=\frac{m'^2}{\pi^4}\chi^2(l) \mathbb{K}^4(l)\\
\mu(l) &=\frac{m'}{\pi} \left[ \mathbb{K}(l)\mathbb{E}(\gamma,l)-\mathbb{E}(l)\mathbb{F}(\gamma,l)   \right]
\end{split}
\ee 

where the shorthand $\chi$ reads

\be
\chi^2(l) = \frac{4\gl_2 m' - 3\gl_2^2 - 2\gl_2 +1}{12} = \frac{l(1-2\gth(l))-3\gth^2(l) +4\gth(l) -1}{3} 
\ee

The weak coupling expansion corresponds to taking $l\to 0$. Expanding $g^2(l)$ and inverting the series one finds a perturbative expression for the elliptic modulus

{\footnotesize
\be
l = \frac{16 \sqrt{2} g}{m'}-\frac{64 g^2}{m'^2}+\frac{164 \sqrt{2} g^3}{m'^3}-\frac{576 g^4}{m'^4}+\frac{695 \sqrt{2} g^5}{m'^5}-\frac{1152 g^6}{m'^6}+\frac{1677 g^7}{\sqrt{2} m'^7}-\frac{1760 g^8}{m'^8}+\frac{13531 g^9}{8\sqrt{2} m'^9}+\cO\left(g^{10}\right)
\ee
}

 that can be plugged into the quantities of interest to obtain a genuine small coupling expansion. Note that, as can be already understood from the first of (\ref{quantities-parametric-form}), the coupling $g$ appears rescaled by the effective mass $m'$ at all orders of perturbation theory. This suggests that, at weak coupling, the width of the eigenvalue distribution $\mu$ must be proportional to the rescaled combination $g=m'\sqrt{\gl}$, as there are no energy scales other that $m'$ in (\ref{saddle-again}). Indeed one finds

\be
\label{mu-weak}
\mu(g) = \sqrt{2} g-\frac{g^3}{\sqrt{2} m'^2}+\frac{15 g^5}{4 \sqrt{2} m'^4}-\frac{165 g^7}{8 \left(\sqrt{2}
   m'^6\right)}+\frac{8555 g^9}{64 \sqrt{2} m'^8}+O\left(g^{11}\right) \qquad g\ll 1
\ee 
 
In order to extract information about the opposite regime one needs to determine the asymptotics for $l\to1$. Since the elliptic integral $\mathbb{K}(l)$ is logarithmically divergent at $l=1$ it is convenient to substitute variables in the following way

\be
l=1-\e^{-L}
\ee

The asymptotics of (\ref{quantities-parametric-form}) as $L\to\infty$ produces 

\be
g^2 = \frac{m'^2 (L-3+4\log(2)) (L+4\log (2))^2}{12 \pi ^4}
\ee

which differs from the result in \cite{Kazakov:1998ji} only in the choice of notation for the logarithmic divergence regulator $L$ (the $\log(2)$ can be reabsorbed in the definition of $L$). After inverting the series one has

\be
L = \tilde g +1 + \frac{1}{\tilde g}+ \cO\left( \tilde g ^{-2} \right)
\ee

being

\be
\tilde g = \frac{\pi^\frac{4}{3}\,2^\frac{2}{3}\,3^\frac{1}{3}\,g^\frac{2}{3}}{m^\frac{2}{3}}
\ee

from which one obtains the strong coupling asymptotics

\be
\label{mu-strong}
\mu(\tilde g) = \frac{m'\tilde g}{2\pi} + \cO\left(\tilde g^{0}\right)
\ee


\subsection*{Momenta of the eigenvalue distribution}

Comparing the large $\gz$ asymptotics (\ref{inverse-map}) and  large $z$ asymptotics of $G(z)$ in (\ref{resolvent-definition}) one can recursively compute even momenta of the eigenvalue distribution $\nu^{(n)}=\int x^{2n}\rho(x)\dd x$ (odd-momenta vanish for parity). In particular for the first few of them one gets 

\be
\label{momenta}
\begin{split}
a_2 &= \frac{m'^2}{g^2}\nu^{(0)}\\
a_3 &= \frac{m'^3-12m'\nu^{(1)}}{8g^4}\\
a_4 &= \frac{5m'^4}{g^4}+\frac{m'^6}{32g^5} + \frac{5m'\nu^{(2)}}{2g^6}-\frac{5m'^3\nu^{(1)}}{4g^6}\\
a_5 &= \frac{m'^7}{128 g^8}+\frac{7 m'^4}{16 g^6}-\frac{35 m'^2 \nu^{(1)}}{4 g^6}-\frac{21 m'^5 \nu^{(1)}}{32 g^8}+\frac{35 m'^3 \nu^{(2)}}{8 g^8}-\frac{7 m' \nu^{(3)}}{2 g^8}\\
\end{split}
\ee

where the $a_k$ have been computed in \cite{Kazakov:1998ji} and read

\be
a_k=\frac{(-1)^{k-1}}{2k-1} \left( \gamma_k+x_3\,\gamma_{k-1} \right) \qquad
\gamma_k= \sum_{p+q+r=k} \left( \begin{array}{c}-\half\\p \end{array}\right)
\left( \begin{array}{c}-\half \\ q \end{array}\right) \left( \begin{array}{c}-\half \\ r \end{array}\right)
x_1^p\, x_2^q\, x_4^r
\ee

As the $x$'s are completely determined in terms of complete elliptic integrals, so are the momenta. For a smooth distribution $\rho(x) = \sum c_2\,x^{2n}$ this is an efficient way to fix the coefficients $c_n$ recursively and compute a polynomial approximation to $\rho(x)$ that converges quickly enough. But this turns out not to be the case of study, at least not for arbitrary values of the coupling constant, as $\rho(x)$ can develop cusps at its endpoints.  The first of equations (\ref{momenta}) is simply the normalization of the eigenvalue distribution; we are in particular interested in the second one because of the fact that the second momentum is proportional to the derivative of the free energy. The expression we get is a slight modification of its homologous in \cite{Kazakov:1998ji} due to the presence of the mass parameter

\be
\label{nu-1}
\nu^{(1)}(l) = \frac{m'^2}{12}-\frac{2m'  \mathbb{K}^2(l) \theta (l)\big[ 5 \theta (l) \big( \theta (l)+l-2 \big)+(l-6) l+6)+(2-l) (l-1)\big]}{5 \pi ^2 \big[ \theta (l) \big( 3 \theta (l)+2 l-4 \big)-l+1 \big]}
\ee

The asymptotics of $\nu^{(1)}$ in the weak/strong coupling phases resembles that of the $m'=1$ case up to a redefinition of the coupling constant $g\to \frac{g}{m}$. Therefore it is legitimate to expect that $\nu^{(1)}(g,m)$ remains a positive and everywhere smooth function.


\section{Analytic continuation and behaviour of the solution at the boundary}
\label{more-solutions}

Interestingly, it turns out that there exist alternative descriptions of the analytic
solution of the master field equation (\ref{saddle-again}). In the limit where the effective mass $m$ is very small compared to average eigenvalue separation, the kernel of   (\ref{saddle-again}) is the discretized version of the Laplace operator acting on the Hilbert kernel

\be
L[H(x)]=\lim_{\gep\to 0} \frac{1}{\gep^2} \left( H(x)-\half H(x-\gep)-\half H(x+\gep)   \right)
\ee

 Setting $m\to 0$ we can take the continuous limit of $L$, so that the saddle point equation can be written as

\be
\label{equation-cubic}
\int_{-\mu}^{\mu} \dd z\,\rho(z) \frac{1}{(x-z)^3} = -\frac{8\pi^2}{m^2\gl}
\ee

Physical values of $Q,M$ imply that $m$ is a real number, which in turn determines boundary conditions to be $\rho(\pm\mu)=0$. Hence we can look for a solution of the kind

\be
\rho_{\scriptsize R}(z) = \sqrt{\mu^2-z^2}\,\ \sum_{n\geq 1} a_n\, z^{2n}
\ee

To lowest order in this expansion, one can fix the first coefficient through the saddle point equation, and
subsequently determine $\mu(\gl)$ requiring normalisation of the density $\int \rho\dd z =1$

\be
a_1=\frac{4\pi}{3m^2 \gl},  \qquad \mu_{\scriptsize R}(\gl) =\sqrt{\frac{m}{\gz \pi}} (6\,\gl)^{1/4} 
\ee

where we have set back dimensionful quantities and $\gz=\sqrt{R_1 R_2}$ was introduced in (\ref{nearly-conformal-case}) and keeps track of the compactification scale (or equivalently the energy scale).
We then have a representation of the solution that behaves near the boundaries according to Wigner semi-circle law

\be
\label{rho-R}
\rho_{\scriptsize R}(z)= \frac{8\pi z^2}{3\mu_{\scriptsize R}^2} \sqrt{\mu_{\scriptsize R}^2-z^2}
\ee

Let us further consider the analytic continuation to imaginary values of $m$. With a slight abuse of notation we set $m^2\to -m^2$. The relevant master field equation then admits solutions with inverse square root behaviour near the boundary of the eigenvalue support

\be
\label{rho-I}
\rho_{\scriptsize I}(z) = \frac{1}{\sqrt{\mu^2-z^2}}\,\ \sum_{n\geq 1} b_n\, z^{2n}
\ee

The lowest term in the polynomial clearly gives null contribution, to first non-trivial order in this expansion one then has

\be
\rho_{\scriptsize I}(z) = \frac{8 z^4}{3\pi \mu_{\scriptsize I}^4 \sqrt{\mu_{\scriptsize I}^2-z^2}} \qquad \mu_{\scriptsize I}(\gl) = \sqrt{\frac{m}{\gz \pi}} (2\,\gl)^{1/4} 
\ee

Through analytic continuation the eigenvalue distribution changes drastically, but it is not a surprising phenomenon. Indeed, changing the sign of $m^2$ amounts to changing the sing of the coupling constant and therefore this is a well known matrix model phase transition. There is a whole variety of questions that arise in this context, such as determining the role of matrix models instantons to this phase transition and whether the transition itself can have relevant effects in our original problem, but we will not discuss them any further herein.



\bibliographystyle{JHEP}
\bibliography{biblio}

\end{document}